\documentclass[twocolumn,aps,pr,superscriptaddress,preprintnumbers,nofootinbib,10pt]{revtex4-2}
\usepackage{amsmath,amssymb}
\usepackage{graphicx}
\usepackage{color}
\usepackage{hyperref}
\usepackage{url}
\usepackage{slashed}
\usepackage{subfigure}
\usepackage[usenames,dvipsnames]{xcolor}
\usepackage{amsmath}
\usepackage{amsfonts}
\usepackage{float} 
\usepackage{amssymb}
\usepackage{euscript}
\usepackage{slashed}
\usepackage[utf8]{inputenc}
\allowdisplaybreaks
\usepackage[normalem]{ulem}
\usepackage{pifont}

\usepackage{dsfont}
\usepackage{MnSymbol}
\usepackage{verbatim}
\usepackage{graphicx}
\usepackage{latexsym}
\usepackage{courier}
\usepackage{xcolor}
\usepackage[normalem]{ulem}
\usepackage{cleveref}

\hypersetup{
colorlinks=true,
citecolor=blue,
citebordercolor=red,
linktoc=all,
linkcolor=blue,
urlcolor=blue
}

\def \pt{\partial}

\def \L{\mathcal{L}}

\def \and{\textmd{and}}

\def \be{\begin{equation}}
\def \ee{\end{equation}}

\def \bea{\begin{eqnarray}}
\def \eea{\end{eqnarray}}

\begin{document}

\title{Three-gluon vertex planar degeneracy: a perturbative explanation}

\author{Daniel O. R. Azevedo}
\affiliation{Instituto de F\'isica, Facultad de Ingenier\'ia, Universidad de la Rep\'ublica, J. H. y Reissig 565, 11000 Montevideo, Uruguay}
\author{Nahuel Barrios} 
\affiliation{Instituto de F\'isica, Facultad de Ingenier\'ia, Universidad de la Rep\'ublica, J. H. y Reissig 565, 11000 Montevideo, Uruguay}
\author{Marcela Pel\'aez}  \affiliation{Instituto de F\'isica, Facultad de Ingenier\'ia, Universidad de la Rep\'ublica, J. H. y Reissig 565, 11000 Montevideo, Uruguay}

\author{Nicol\'as Wschebor}  \affiliation{Instituto de F\'isica, Facultad de Ingenier\'ia, Universidad de la Rep\'ublica, J. H. y Reissig 565, 11000 Montevideo, Uruguay}

\begin{abstract}
We provide a perturbative explanation for the planar degeneracy of the three-gluon vertex in Yang-Mills theory, using the Curci-Ferrari (CF) model in the Landau gauge. By employing an infrared safe renormalization scheme whose coupling remains regular at all scales, we perform a one-loop calculation for arbitrary kinematics in both three and four dimensions. Our results demonstrate that the observed planar degeneracy can be understood naturally due to perturbative reasons: i) the tree-level form factor dominates the vertex structure simply because perturbative corrections are suppressed; ii) the one loop corrections to this form factor remain remarkably flat across the allowed phase space. We quantify the deviations from the symmetric configuration, finding that the mean error of the symmetric approximation, in the case of the tree-level form factor, peaks at approximately 3\% both in three and four dimensions, with the most significant departures occurring near soft-gluon configurations due to tamed ultraviolet collinear divergences. This work explains previous lattice and Dyson-Schwinger studies while reinforcing the capacity of the mass-extended Faddeev-Popov Lagrangian to capture complex infrared QCD dynamics through perturbative analytical techniques.
\end{abstract}

\maketitle

\section{Introduction \label{Sec:Intro}}

Understanding the infrared (IR) regime of Quantum Chromodynamics (QCD) is, still today, one of the major challenges of particle physics. In particular, this is true for the mechanisms behind color confinement and the spontaneous breaking of chiral symmetry. The former is associated with the formation of the hadronic spectrum, as it precludes the existence of free quarks and gluons as asymptotic states, whereas the latter is the main responsible for the generation of most of hadron masses. 

One of the most salient features of QCD is asymptotic freedom \cite{Politzer:1973fx,Gross:1973id}, which ensures that the gauge coupling becomes small at sufficiently high energies, thus enabling perturbation theory to be used reliably. However, the presence of a Landau pole in the IR regime of standard perturbation theory prevents its extension to low momenta, making it necessary to employ alternative methods to access this regime.

In this pursuit, lattice simulations stand out as a first-principles, fully non-perturbative approach. In the last decades, these simulations have, for instance, successfully computed the spectrum of hadronic masses, as well as their constituent quark masses in agreement with experimental data, attesting to the validity and predictive power of the method \cite{EuropeanTwistedMass:2008pab,Fodor:2012gf,EuropeanTwistedMass:2014osg,Bazavov:2017dsy,Bussone:2023kag}. In addition, lattice results serve as benchmarks for the study of correlation functions in both pure Yang-Mills (YM) theory \cite{Mandula:1987rh,Catumba:2021yly,Bonnet:2000kw,Cucchieri:2006tf,Cucchieri:2007rg,Cucchieri:2008qm,Duarte:2016ieu,Duarte:2016iko,Maas:2020zjp,Colaco:2024gmt,Oliveira:2012eh,Iritani:2009mp,Bornyakov:2009ug,Bogolubsky:2009dc,Maas:2007uv} and QCD \cite{Kizilersu:2021jen,Silva:2010vx,Bowman:2004jm,Marques:2023cmi}. These objects are crucial for continuum approaches, such as Dyson-Schwinger equations (DSE) \cite{vonSmekal:1997ohs,Alkofer:2000wg,Fischer:2002hna,Zwanziger:2001kw,Maas:2004se,Aguilar:2007ie,Aguilar:2008xm,Boucaud:2008ji,Boucaud:2008ky,Huber:2007kc,Huber:2012kd,Huber:2016tvc,Huber:2018ned,Huber:2020keu,Fischer:2008uz,Fischer:2020xnb,Cyrol:2016tym} and those based on the functional renormalization group (FRG) \cite{Ellwanger:1995qf,Pawlowski:2003hq,Fischer:2004uk,Fischer:2006vf,Dupuis:2020fhh}, that typically need to determine a set of correlation functions in order to evaluate observables.

   
A different approach to the IR of QCD comes from the observation that not only naive perturbation theory fails in the IR, but that the Faddeev-Popov (FP) gauge fixing procedure itself is no longer valid at such energy scales. Contrary to the lattice approach, any continuum method needs to work in a specific gauge, commonly introduced via the FP procedure \cite{Faddeev:1967fc}. In particular, typical gauge-fixing conditions do not uniquely select a representative of each gauge orbit, as multiple gauge-equivalent field configurations ("copies") satisfy the same condition, as observed first by Gribov \cite{Gribov:1977wm}. A possible way to tackle this issue involves restricting the functional integration over the gauge fields to a region without Gribov copies. This program has been partially accomplished by the (refined) Gribov-Zwanziger (GZ) approach \cite{Gribov:1977wm,Zwanziger:1988jt,Zwanziger:1989mf,Dudal:2008sp,Vandersickel:2012tz}, which is capable of removing infinitesimal copies but not in a completely first-principles setting. However, the complete elimination of such Gribov copies has proven to be a formidable task, remaining elusive for decades. 

An alternative route to describe the infrared dynamics of QCD, central for this article, is based on phenomenological grounds. It takes inspiration from the so called decoupling solution of the gluon propagator observed in lattice simulations in Landau gauge in dimension greater than two \cite{Cucchieri:2007rg,Bornyakov:2009ug,Bogolubsky:2009dc,Maas:2011se,Oliveira:2012eh,Duarte:2016iko}, which corresponds to gluon propagator saturating to a finite value at vanishing momentum, similarly to a massive field. This motivates a phenomenological approach by adding a gluon mass to the standard FP Lagrangian \cite{Tissier:2010ts,Tissier:2011ey}. Such Lagrangian corresponds to a particular case of the Curci-Ferrari (CF) Lagrangian\footnote{Another way to introduce the gluon mass is by the screened mass expansion \cite{Siringo:2015wtx,Siringo:2016jrc,Comitini:2021kxj}. It differs from the CF approach due to the mass term being summed and subtracted, leaving the original Lagrangian unmodified, and performing a double expansion, which induces a mass term at any finite perturbative order.}\cite{Curci:1976bt}, restricted to the Landau gauge and is renormalizable. Even though there are various mechanisms that could eventually explain the emergence of this term, see, \textit{e.g.} \cite{Verschelde:2001ia,Serreau:2012cg,Aguilar:2019kxz,Ferreira:2024czk,Ferreira:2025anh}, there is still no unanimous consensus on this matter. The philosophy underlying the use of the CF model is to explore the consequences of such a term rather than to investigate its origin from first principles.

The first quantities calculated by employing the CF model were two- and three-point correlation functions in Landau gauge, since these objects have already been computed by lattice simulations and, as a result, allowed for testing the validity and accuracy of the model. These calculations were performed perturbatively for all momentum scales, displaying, in general terms, a very good agreement with lattice data already at one-loop order \cite{Tissier:2010ts,Tissier:2011ey,Pelaez:2013cpa,Pelaez:2014mxa,Pelaez:2015tba}, with a few-percent level of error once two-loop corrections were included \cite{Gracey:2019xom,Barrios:2020ubx,Barrios:2021cks,Barrios:2024ixj}.

The use of perturbative techniques seems, at first, surprising. However, it is well justified, in the first place, from the results of lattice simulations for the gauge coupling that can be extracted from various kinds of vertices in Landau gauge. These show no sign of Landau pole in the IR. Instead, the various possible definitions of the coupling remains moderate at all scales, reaching a maximum around 0.5 GeV \cite{Bonnet:2000kw,Bogolubsky:2009dc,Bornyakov:2009ug,Iritani:2009mp,Maas:2011se,Oliveira:2012eh,Boucaud:2011ug,Bogolubsky:2009dc,Duarte:2016iko} and decreasing towards the deep infrared. Furthermore, this behavior corresponds to an expansion parameter for perturbation theory of $\lambda=g^2 N_c/(16\pi^2)\lesssim 0.3$, where $g$ is the gauge coupling \cite{Weinberg:1995mt}. This result suggests that some sort of perturbation theory should work, as the error of a one-loop evaluation would be of the order $(0.3)^2\simeq 10\%$. This is in agreement with one-loop calculations of correlation functions in the CF model, which show this level of error when compared with results from lattice simulations \cite{Tissier:2011ey,Pelaez:2013cpa,Pelaez:2014mxa,Pelaez:2015tba}. In a similar way, two-loop calculations give typical error bars of the order of $4 \%$ \cite{Gracey:2019xom,Barrios:2020ubx,Barrios:2021cks,Barrios:2024ixj}.

Besides correlation functions, the model has successfully been employed to describe some aspects of the phase structure of YM/QCD \cite{MariSurkau:2026irs,Surkau:2024zfb,Maelger:2019cbk,Maelger:2018vow}, spontaneous chiral symmetry breaking \cite{Pelaez:2020ups} and, more recently, confinement \cite{Pelaez:2024mtq}. The model has also been used to compute the pion decay constant \cite{Pelaez:2022rwx}, heavy mesons spectra \cite{Alvez:2025wek}, parton distribution and fragmentation functions \cite{Bopsin:2025vhz} and it has been extended to Minkowski space-time \cite{Oribe:2025ezp}. Beyond that, it leads to some interesting neutron star phenomenology \cite{Song:2019qoh,Suenaga:2019jjv,Kojo:2021knn}. These results demonstrate that, with a minimal modeling, the CF model captures the infrared dynamics while remaining, to a great extent, analytically tractable. However it must be stressed that, when considering light quarks, the treatment requires to go beyond strict perturbation theory as, in that case, some couplings become too large to be treated perturbatively \cite{Pelaez:2017bhh,Pelaez:2021tpq,Pelaez:2022rwx}.

As stated before, a key advantage of the CF for pure YM model (or QCD with heavy quarks) is that correlation functions are obtained self-consistently from the Lagrangian by using standard perturbative techniques. This simplicity is particularly powerful. For instance, it allowed the CF framework to be one of the first approaches to compute the three-gluon vertex for arbitrary kinematics\cite{Pelaez:2013cpa}, despite the complex tensor structures involved. The results show excellent agreement with lattice data, demonstrating that the model captures the infrared dynamics while remaining analytically tractable.

In this work, we concentrate on the three-gluon vertex, specifically on a property of this quantity known as planar degeneracy \cite{Pinto-Gomez:2022brg,Aguilar:2023qqd,Pinto-Gomez:2024mrk}. The three-gluon vertex has already been evaluated within the CF framework, at one-loop order for arbitrary kinematics \cite{Pelaez:2013cpa} and at two-loop in the soft-gluon configuration \cite{Barrios:2022hzr}, displaying very good agreement with lattice data. This quantity has attracted significant interest in the SDE community, partly due to its central role in the interpretation that the dynamical generation of the gluon mass occurs via the Schwinger mechanism \cite{Aguilar:2019kxz,Aguilar:2022thg,Aguilar:2023mdv,Ferreira:2024czk,Ferreira:2025anh}.

More recently, particular attention has been given to the planar degeneracy of the three-gluon vertex \cite{Pinto-Gomez:2022brg,Aguilar:2023qqd,Pinto-Gomez:2024mrk}, observed in lattice simulations and previously hinted in DSE investigations \cite{Eichmann:2014xya}. Briefly, it states that the transversely projected three-gluon vertex can be accurately described by using only a symmetric combination of the external momenta $s^2=(p^2+r^2+k^2)/2$. Therefore, configurations lying on the same $s^2$ plane share nearly identical form factors, hence the name planar degeneracy. Beyond that, lattice and DSE studies show that the tree level form factor dominates over the others, meaning the vertex can be approximated by its classical structure for general kinematics. This nice and simple property looks surprising for lattice and DSE analysis, where no small parameter is assumed. We will show in the present article that it can be very well understood from a purely simply perturbative analysis. This gives a further support to the validity of perturbation theory in pure YM theory in the IR. In fact, a particular case of the one-loop analysis (the bisectoral configuration for the tree-level like form factor) was already considered in a previous work \cite{Pinto-Gomez:2022brg}. In the present work we generalize that analysis to arbitrary configuration of momenta and incorporating Renormalization-Group (RG) effects that are always important in a model where the coupling is moderate but not very small. This can only be done properly by using an appropriate renormalization scheme that avoids the presence of an IR Landau pole and in the context of a renormalizable model as the CF.


Accordingly, in this article, we investigate the planar degeneracy of the three-gluon vertex within the Curci-Ferrari model by employing a one-loop approximation, which allows us to extract analytical expressions for all the form factors of the vertex and, as a result, show how this property emerges in the IR. The outline of the article is the following: in \cref{Sec:CF}, we give a general review of the Curci-Ferrari model, discussing its renormalization scheme and renormalization group properties. In \cref{Sec:3gluonvert}, we present the generalities of the three-gluon vertex, the choice of tensor basis and kinematical configuration. In \cref{Sec:planar}, we discuss how the planar degeneracy appears in the CF model and present the results for 3 and 4 dimensions. At the end, in \cref{Sec:conclusions}, we present the conclusions of this work.

\section{The Curci-Ferrari model\label{Sec:CF}}

We define the Curci-Ferrari model in the Landau gauge through the following Lagrangian density:
\be
\L = \frac{1}{4}F^a_{\mu\nu}F^a_{\mu\nu}+i h^a\pt_\mu A^a_\mu + \pt_\mu\Bar{c}^a(D_\mu c)^a + \frac{1}{2}m^2 A^a_\mu A^a_\mu,
\ee
where $D^{ab}_\mu = \delta^{ab}\pt_\mu - gf^{abc}A^c_\mu$ is the covariant derivative in the adjoint representation of the gauge group $SU(N)$, $f^{abc}$ are the structure constants of $SU(N)$ and $g$ and $m$ are the bare gauge coupling and gluon mass, respectively. The field strength is given by $F^a_{\mu\nu}=\pt_\mu A^a_\nu - \pt_\nu A^a_\mu + gf^{abc}A^b_\mu A^c_\nu$, $c,\bar{c}$ are the Faddeev-Popov ghost fields and $h$ is the Nakanishi-Lautrup auxiliary field, which enforces the Landau gauge condition $\pt_\mu A^a_\mu = 0$. The model is renormalizable in $d\leq 4$ dimensions and reduces to the standard YM theory in the UV, where $m$ is negligible. The presence of a gluon mass term, introduced at the level of the gauge-fixed theory, regularizes the IR behavior of the model, while being a minimal modification to the FP Lagrangian \cite{Pelaez:2021tpq}, but it breaks the BRST symmetry of the Faddeev-Popov Lagrangian. Instead, it satisfies a modified, non-nilpotent BRST symmetry, which can be used to prove its renormalizability \cite{deBoer:1995dh}. 

The divergences appearing at one-loop order and beyond can be regularized through dimensional regularization and reabsorbed into the fields and parameters of the model with the introduction of the renormalization factors. These quantities relate the initial bare fields and parameters of the Lagrangian with their renormalized ones as follows
\bea
&A_B^{\mu,a} = Z_A^{1/2} A_R^{\mu,a}, \quad c_B^{a} = Z_c^{1/2} c_R^{a}, \quad \Bar{c}_B^{a} = Z_c^{1/2} \Bar{c}_R^{a},\nonumber
\\
&g_B = Z_g g_R, \quad m^2_B = Z_{m^2}m^2_R.
\eea

The renormalization factors are fixed by a given renormalization scheme, which imposes conditions over some renormalized propagators and vertices at a certain renormalization scale $\mu$. Remarkably, the CF model allows for renormalization schemes which are Landau pole free, \textit{i.e.}, regular at all scales \cite{Reinosa:2017qtf}. One of such schemes is the so-called infrared safe (IRS) scheme \cite{Tissier:2011ey} defined by the following renormalization conditions
\be
D^{-1}_A (p=\mu) = \mu^2 + m^2(\mu), \quad D^{-1}_c (p=\mu) = \mu^2, 
\ee
\be\label{nonrencond}
Z_gZ_A^{1/2}Z_c = 1, \quad Z_{m^2}Z_AZ_c=1,
\ee
where $D_A$ and $D_c$ refer to the gluon and ghost propagators, respectively. The conditions \eqref{nonrencond} are to be taken as extensions of two non-renormalization theorems to include the finite parts of the renormalization factors as well. The first identity is the Taylor scheme \cite{Taylor:1971ff}, which emerges from the fact that the ghost-gluon vertex does not receive loop corrections in the limit of vanishing ghost momentum \cite{Taylor:1971ff}, whereas the second one is a non-renormalization theorem for the gluon mass term \cite{Gracey:2002yt,Dudal:2002pq,Wschebor:2007vh,Tissier:2008nw}. 

Besides renormalization, when performing perturbative computations we need to take into account that, in general, an $n$-loop expansion introduces logarithms of the form $\ln p^2/\mu^2$  (and up to $(\ln p^2/\mu^2)^n$) in the expression of a general correlation function. This could invalidate the use of the perturbative framework, since for $p\gg \mu$ all loop contributions are of the same order. This issue can be fixed with the introduction of a running scale $\mu(p)$, such that $\mu(p)\approx p$. 

However, in order to compare with lattice data we need to choose a fixed renormalization scale $\mu_0$. To make amends for this issue, one can use the Callan-Symanzik equation, which relates the same correlation function evaluated at two different renormalization scales, such as $\mu_0$ and $\mu(p)$. For a purely gluonic vertex function with $n_A$ external gluon legs, this equation reads
\be\label{CSeq}
\left(\mu\pt_\mu -\frac{1}{2}n_A\gamma_A +\beta_g\pt_g +\beta_{m^2}\pt_{m^2} \right)\Gamma^{(n_A)}=0,
\ee
where we introduced the $\beta$-functions and anomalous dimension $\gamma_A$ 
\be
\beta_X(g,m^2)= \mu \left.\frac{\pt X}{\pt \mu}\right|_{g_B,m^2_B}, 
\ee
\be
\gamma_A(g,m^2)= \mu \left.\frac{\pt \ln Z_A}{\pt \mu}\right|_{g_B,m^2_B},
\ee
with $X=\{g,m^2\}$. The solution for the Callan-Symanzik equation \eqref{CSeq} can be written as
\begin{multline}\label{CSresult}
\Gamma^{(n_A)}(\{p_i\},\mu,g(\mu),m^2(\mu))\\= z_A(\mu;\mu_0)^{n_A/2}\Gamma^{(n_A)}(\{p_i\},\mu_0,g_0,m^2_0) ,
\end{multline}
where
\be\label{zscale}
z_A(\mu;\mu_0)= \exp\left[\int_{\mu_0}^{\mu}\frac{d\mu'}{\mu'}\gamma_A(g(\mu'),m^2(\mu'))\right]. 
\ee 

The advantage of using relation \eqref{CSresult} is that it allows us to compute $\Gamma^{(n_A)}(\mu_0)$ perturbatively at any given scale, even those where $p\gg\mu_0$. By using the conditions \eqref{nonrencond} and the independence of the bare quantities from the renormalization scale, one can obtain a simple expression for $\gamma_A$ in terms of the $\beta$-functions:
\be\label{gammaA}
\gamma_A = 2\left(\frac{\beta_{m^2}}{m^2} -\frac{\beta_g}{g}\right).
\ee
By inserting \eqref{gammaA} in the integral \eqref{zscale}, we get an analytical expression for the factor $z_A(\mu;\mu_0)$ in terms of the running gluon mass and the gauge coupling, given by
\be
z_A(\mu;\mu_0) = \frac{m^4(\mu)}{m^4(\mu_0)}\frac{g^2(\mu_0)}{g^2(\mu)}.
\ee
This feature is another advantage of working with the IRS scheme. The solution  \eqref{CSresult} of the Callan-Symanzik equation, relating the vertex at different scales, is then given by
\bea
& \Gamma^{(n_A)}(\mu_0,g(\mu_0),m^2(\mu_0)) = \\ \nonumber
& \left(\frac{m^4(\mu)}{m^4(\mu_0)}\frac{g^2(\mu_0)}{g^2(\mu)}\right)^{-n_A/2}\Gamma^{(n_A)}(\mu,g(\mu),m^2(\mu)).
\eea
The parameters $g(\mu)$ and $m^2(\mu)$ can be obtained by integrating their respective $\beta$ functions with initial conditions, $g_0$ and $m^2_0$, at the scale $\mu_0$. These initial conditions can be chosen so as to to fit the lattice data for the gluon and ghost two-point functions. This presents two main advantages: the lattice data for two point functions is far more precise than for higher correlation functions, allowing for a higher accuracy and, with the parameters determined from the propagators, the results for the interaction vertices become a prediction of the model, serving as a an important benchmark for the precision of the model. 

In what follows, we adopt the prescription $\mu=p$, which reproduces the desired UV behavior to avoid large logarithms as presented in \cite{Pelaez:2013cpa}.

\section {The transversely projected three-gluon vertex at 1-loop \label{Sec:3gluonvert}}

The color structure of the 1PI three-gluon vertex is exactly given by the structure constant $f^{abc}$ of the gauge group $SU(N)$ \cite{Smolyakov:1980wq}. Factoring out the gauge coupling, we can write the vertex as
\be\label{3gluon}
\Gamma^{(3)}_{A^a_\mu A^b_\nu A^c_\rho}(p,r,k) = -i g f^{abc} \Gamma_{\mu\nu\rho}(p,r,k).
\ee 

As for the Lorentz structure, there are 14 terms that can be constructed from combinations of the external momenta and the metric. But since the vertex is symmetric under the exchange of any two external gluon legs, only six of those tensors have independent form factors. One way to arrange the vertex in view of this symmetry is following the Ball-Chiu (BC) decomposition \cite{Ball:1980ax}:
\begin{widetext}
\bea\label{BCdec}
&\Gamma_{\mu\nu\rho}(p,r,k) &= A(p^2,r^2,k^2)\delta_{\mu\nu}(p-r)_\rho + B(p^2,r^2,k^2)\delta_{\mu\nu}(p+r)_\rho - C(p^2,r^2,k^2)(\delta_{\mu\nu}p\cdot r - p_\nu r_\mu)(p-r)_\rho\nonumber\\
&&+\frac{1}{3}S(p^2,r^2,k^2)(p_\rho r_\mu k_\nu + p_\nu r_\rho k_\mu) + F(p^2,r^2,k^2)(\delta_{\mu\nu}p\cdot r - p_\nu r_\mu)(p_\rho r\cdot k  - r_\rho p\cdot k  )\\
&&+H(p^2,r^2,k^2)\left[ -\delta_{\mu\nu}(p_\rho r\cdot k  - r_\rho p\cdot k  )+ \frac{1}{3}(p_\rho r_\mu k_\nu - p_\nu r_\rho k_\mu) \right] + \text{cyclic permutations}\nonumber
\eea
\end{widetext}

The scalar functions have the following symmetry properties: A, C and F are symmetric under permutation of the first two arguments; B is antisymmetric under permutation of the first two arguments; H is completely symmetric and S is completely antisymmetric. This decomposition concerns 1PI functions. However, these are not directly accessible to lattice simulations, which can only compute connected correlation functions, meaning that all external legs are contracted with full gluon propagators. As a consequence, the longitudinally coupled scalar coefficients, such as the B and S functions in the BC decomposition \eqref{BCdec}, cannot be obtained through lattice computations in the Landau gauge.

Consequently, in order to compare with lattice data, we focus on the transversely projected three-gluon vertex, that matches what can be observed on the lattice. This is,
\be\label{transvprojvert}
\Bar{\Gamma}_{\mu\nu\rho}(p,r,k) = P^\perp_{\mu\mu'}(p)P^\perp_{\nu\nu'}(r)P^\perp_{\rho\rho'}(k)\Gamma_{\mu'\nu'\rho'}(p,r,k),
\ee
where $P^\perp_{\mu\nu}(p) = \left(\delta_{\mu\nu}-\frac{p_\mu p_\nu}{p^2}\right)$ is the usual transverse projector. Since the transverse projectors remove the longitudinal components of the vertex, $\Bar{\Gamma}$ can be written in terms of a four-dimensional basis of transverse tensors for a general kinematic configuration,
\be\label{generalbasis}
\bar{\Gamma}_{\mu\nu\rho}(p,r,k) = \sum_{i=1}^4 \bar{\Gamma}_i(p^2,r^2,k^2)\tau^i_{\mu\nu\rho}(p,r,k),
\ee
where $\tau^i_{\mu\nu\rho}(p,r,k)$ is an element of the transverse basis and $\bar{\Gamma}_i(p^2,r^2,k^2)$ is the corresponding form factor.

\begin{figure}
    \centering
    \includegraphics[width=0.3\linewidth]{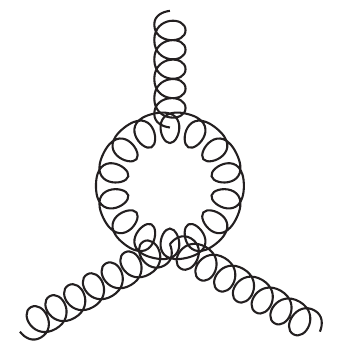}\hfill
    \includegraphics[width=0.4\linewidth]{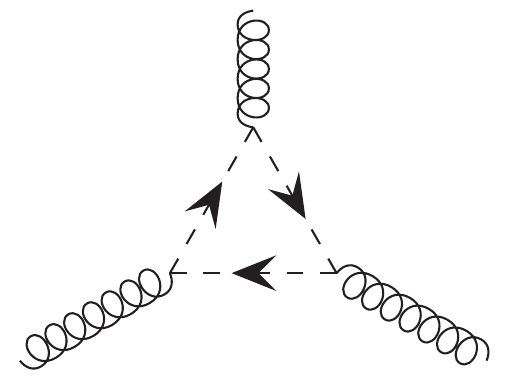}
    \includegraphics[width=0.4\linewidth]{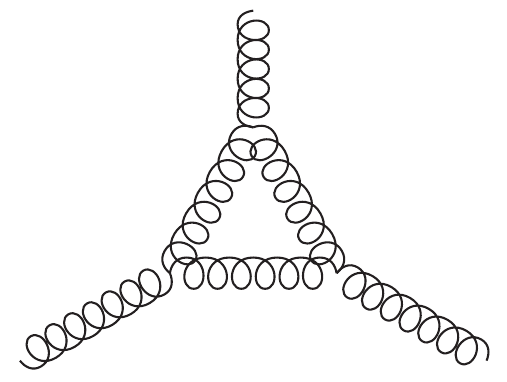}
    \caption{Feynman diagrams appearing at 1-loop order for the three-gluon vertex.}
    \label{feyndiag}
\end{figure}

At one-loop order, there are three Feynman diagrams contributing to the three-gluon vertex, see \cref{feyndiag}. Concerning the singular behavior and the zero-crossing of the three-gluon vertex, it can be shown that, in the CF model, these stem from an effective ghost loop diagram at one-loop order with dressed ghost-gluon vertices, at any order of the perturbative expansion \cite{Barrios:2022hzr}. At one-loop order, this is simply the ghost loop diagram. Since the only tensor structure coming from this diagram corresponds to the tree-level vertex, its associated form factor is the only one divergent in the limit of vanishing momenta, whereas the remaining three are regular in this limit. All the considerations made in \cref{Sec:CF} for a general gluonic $n_A$-point function, in particular regarding the renormalization group, apply here for $n_A=3$.


\subsection{Tensor basis\label{basis}}

As stated before, the three-gluon vertex \eqref{3gluon} is completely symmetric under Bose symmetry, that is, the exchange of any two gluon legs including the corresponding color and Lorentz indices. By factoring out the structure constants $f^{abc}$, we are thus left with a vertex which is antisymmetric under the simultaneous exchange of any two indices and the corresponding momenta. Therefore, a suitable choice of tensor basis $\tau^i_{\mu\nu\rho}(p,r,k)$ is one which encompasses this antisymmetric property, so that the form factors $\bar{\Gamma}_i(p^2,r^2,k^2)$ are invariant under any permutation of the external momenta. Moreover, by requiring the tensors to be of mass dimension one, like the tree-level expression, all form factors are dimensionless and can be directly compared to one another \cite{Aguilar:2023qqd}. 

The presence of the infrared divergence when all momenta vanish, together with the perturbative limit of the CF model, impose that the transversely projected tree-level tensor, defined as
\be
\bar{\Gamma}_{\mu\nu\rho}^0(p,r,k) = P^\perp_{\mu\mu'}(p)P^\perp_{\nu\nu'}(r)P^\perp_{\rho\rho'}(k)\Gamma^0_{\mu'\nu'\rho'}(p,r,k),
\ee
where
\be
\Gamma^0_{\mu\nu\rho}(p,r,k)=(k-r)_\mu \delta_{\nu \rho}+(r-p)_\nu \delta_{\mu \rho}+(p-k)_\rho \delta_{\mu \nu},
\ee
should be taken as one of the elements of the four-dimensional transverse basis $\tau^i_{\mu\nu\rho}(p,r,k)$, so that it fully encapsulates the infrared divergence mentioned above in its form factor $\bar{\Gamma}_0(p^2,r^2,k^2)$. Moreover, as $\Gamma^0_{\mu\nu\rho}$ is the only tensor to appear in the classical Lagrangian, it also contains the divergences in the one loop Feynman diagrams, due to the multiplicative renormalizability of the Curci-Ferrari model \cite{Curci:1976bt,deBoer:1995dh,Delduc:1989uc}. 

A suitable choice of basis is the one from \cite{Aguilar:2023qqd}\footnote{To avoid cluttering, we omit the dependence on the momenta of the tensors.}:
\bea\label{basisaguilar}
&\tau^1_{\mu\nu\rho} &= \bar{\Gamma}_{\mu\nu\rho}^0 ,\nonumber\\
&\tau^2_{\mu\nu\rho} &= \frac{3}{2s^2}(p-r)_{\mu'}(r-k)_{\nu'}(k-p)_{\rho'}\times\\
&&P^\perp_{\mu\mu'}(k)P^\perp_{\nu\nu'}(p)P^\perp_{\rho\rho'}(r),\nonumber\\
&\tau^3_{\mu\nu\rho} &= -\frac{\tau^2_{\mu\nu\rho} }{4}-3\frac{t^4_{\mu\nu\rho}}{s^2},\\
&\tau^4_{\mu\nu\rho} &=\left(\frac{3}{2s^2}\right)^2 (t^1_{\mu\nu\rho} + t^2_{\mu\nu\rho} + t^3_{\mu\nu\rho})\nonumber,
\eea
where $s^2=(p^2+r^2+k^2)/2$ and $t^i$ stands for the transverse tensors of the Ball-Chiu basis \cite{Ball:1980ax,Aguilar:2019jsj}, 

\bea\label{eq:def_ts}
&t_1^{\mu\nu\rho}&=[(p\cdot r)\delta^{\mu\nu}-p^\nu r^\mu][(r\cdot k)p^\rho-(p\cdot k)r^\rho],\\ \nonumber 
&t_2^{\mu\nu\rho}&=[(r\cdot k )\delta^{\nu \rho}-r^\rho k^\nu][(p \cdot k)r^\mu -(r\cdot p)k^\mu],\\ \nonumber
&t_3^{\mu\nu\rho}&=[(p \cdot k)\delta^{\mu \rho}-k^\mu p^\rho][(p\cdot r)k^\nu-(r\cdot k)p^\nu],\\ \nonumber
&t_4^{\mu \nu \rho}&=\delta^{\nu \rho}[(r \cdot p)k^\mu-(p\cdot k)r^\mu]\\ \nonumber
&&+\delta^{\mu \rho}[(r\cdot k)p^\nu-(p\cdot r)k^\nu]+p^\rho r^\mu k^\nu\\ \nonumber
&&+\delta^{\mu \nu}[(p\cdot k)r^\rho-(r\cdot k)p^\rho]-p^\nu r^\rho k^\mu,
\eea
which is a minimal modification of the basis $\{\tilde{\lambda}^i_{\mu\nu\rho}\}$ employed in the lattice studies \cite{Pinto-Gomez:2022brg,Pinto-Gomez:2024mrk}, differing only in the definition of the element $\tau^3_{\mu\nu\rho}$ through the relation
\be\label{latticebasis}
\tau^3_{\mu\nu\rho} = \tilde{\lambda}^3_{\mu\nu\rho} - \frac{3}{2}\tilde{\lambda}^1_{\mu\nu\rho}.
\ee

For some special kinematic configurations, the transversely projected vertex tensor space has its dimension reduced: for bisectoral configurations, where two of the momenta have the same magnitude and form the same angle with the third (e.g. $p^2=r^2$ and $\theta_{pk}=\theta_{rk}$), it is three-dimensional; for symmetric configurations, where all momenta have the same magnitude ($p^2=r^2=k^2$), it is two-dimensional; for soft configurations, where one of the momenta is zero (e.g. $p=0$), it is one-dimensional. The tree-level tensor $\bar{\Gamma}_{\mu\nu\rho}^0$ does not vanish for any particular kinematical configuration, meaning the other three tensors of the basis either vanish or become linearly dependent with $\bar{\Gamma}_{\mu\nu\rho}^0$ in those limits. In an ideal basis, the tensor elements would exactly vanish at these kinematical configurations, so there would be no linear combination of the form factors in any kinematics, so that they could be read directly from $\bar{\Gamma}_i(p^2,r^2,k^2)$. For the basis defined \eqref{basisaguilar}, only the tree-level tensor is non-vanishing in the soft-gluon limit, so the form factor comes exactly from $\tau^1_{\mu\nu\rho}$. This is the reason behind the different choice of basis from the one used in lattice studies (see Eq. \eqref{latticebasis}), where the soft-gluon form factor is given by a linear combination of the $\tilde{\lambda}^1_{\mu\nu\rho}$ and $\tilde{\lambda}^3_{\mu\nu\rho}$ form factors.

For later convenience, we state that, in the symmetric configuration, the form factor associated with the tree-level tensor component $\bar{\Gamma}_{\mu\nu\rho}^0$ is defined as $\Gamma_1^{sym}(p^2)=\bar{\Gamma}_1(p^2,p^2,p^2)-\bar{\Gamma}_3(p^2,p^2,p^2)+\frac{\bar{\Gamma}_4(p^2,p^2,p^2)}{4}$.

\subsection{Kinematics\label{Sec:kinematics}}

The three gluon vertex is constrained by momentum conservation ($p+r+k=0$), which implies its dependence on three independent kinematical scalar variables. These can be expressed as the squared momenta $p^2,r^2,k^2$. Equivalently, two of the momenta can be replaced by the angle between any two momenta (e.g. $\cos \theta_{pr}=(k^2-p^2-r^2)/({2}\sqrt{p^2r^2})$
). As previously discussed, the form factors depend only on Bose symmetric combinations of these variables, like the sum of the squared momenta $s^2=(p^2+r^2+k^2)/2$, since the chosen tensor basis fully captures the antisymmetric property of the vertex. 

\begin{figure}
    \centering
    \includegraphics[width=\linewidth]{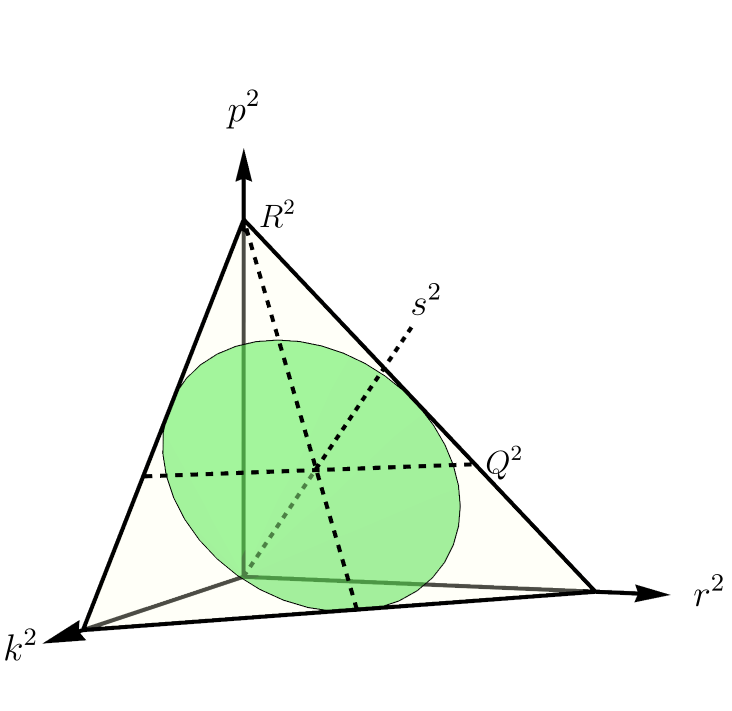}
    \caption{Kinematical configuration of the three gluon vertex. The light green disk represents a slice of the cone centered around the $p^2=r^2=k^2$ line for a fixed value of $s^2$ and it contains all the allowed configurations of the vertex.}
    \label{kinematics}
\end{figure}

Nevertheless, to explore the planar degeneracy approximation, it is useful to express the form factors in terms of asymmetrical variables. This can be achieved by restricting the kinematical configuration to the plane $s^2=s_0^2$ within the first octant of the ($p^2,r^2,k^2$) space. This plane forms an equilateral triangle perpendicular to the diagonal of the octant, with vertices at $p^2=s^2$, $r^2=s^2$ and $k^2=s^2$ and edges where one the squared momenta vanishes. Introducing the coordinates $Q^2$ and $R^2$ in this plane \cite{Eichmann:2014xya,Pinto-Gomez:2022brg} 
\bea\label{asymcoord}
Q^2 = \frac{r^2-k^2}{2}, \quad  R^2=\frac{2p^2-r^2-k^2}{6},
\eea 
the allowed kinematics map onto a disk centered around the symmetric point $p^2=r^2=k^2$ and inscribed in the $s_0^2$ triangle, with its radius proportional to the Bose symmetric variable $s^2$ (see Fig. \ref{kinematics}). The boundary of the disk corresponds to all possible collinear configurations, including the three soft-gluon limits where the circle touches the edges of the $s_0^2$ triangle. Consequently, the full space of kinematical configurations form a cone in the first octant, centered around the $p^2=r^2=k^2$ line. 

The special kinematical configurations discussed in section \ref{basis}, where the dimensionality of the tensor basis reduces, are also neatly captured in this geometrical picture. In this representation, bisectoral configurations, such as $p^2=r^2\neq k^2$, appear as radial lines coming from the intersection of the $s_0^2$ triangle and the inscribed disk; the symmetric and soft gluon configurations correspond to specific points in the disk: the symmetric point lies in the center, while soft-gluon limits are located where the disk touches the edges of the triangle at each bisectoral line. This visualization is useful for examining how each form factor varies across different configurations, given a fixed $s^2$.

It is important to note that only soft-gluon configurations are entirely defined by a single form factor, $\bar{\Gamma}_1(0,k^2,k^2)$, in the basis \eqref{basisaguilar}. For bisectoral and symmetric configurations, the corresponding form factors are instead given by linear combinations of $\bar{\Gamma}_i(p^2,r^2,k^2)$ evaluated at the appropriate limits; see Ref. \cite{Pinto-Gomez:2024mrk} for details.

In what follows, we will use the coordinate system defined in Eq.~\eqref{asymcoord} to study how the three-gluon vertex varies across general configurations relative to the symmetric and soft-gluon ones, elucidating why and to what extent the planar degeneracy approximation holds. 

\begin{figure}
    \centering
    \includegraphics[width=\linewidth]{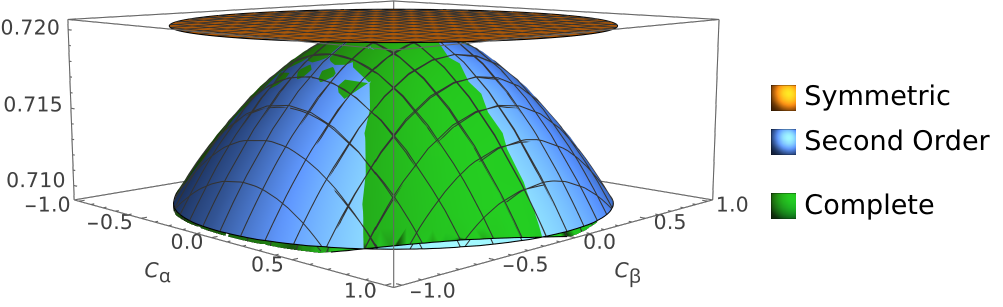}
    \caption{Graphical representation of the three dimensional form factor $\bar{\Gamma}_1(p^2,r^2,k^2)$ The coordinates $Q^2$ and $R^2$ are normalized as $C_\alpha$ and $C_\beta$ (see Eq. \eqref{eq:defs_ca_cb}), such that their range is $[-1,1]$. The flat upper disc represents the zeroth order symmetric approximation and would correspond to exact planar degeneracy, while the overlapping surfaces are the second order expansion and the complete form factor expression. We fix $s^2=2.25$ GeV$^2$.}
    \label{desarrollo3d}
\end{figure}
\begin{figure}
    \centering
    \includegraphics[width=\linewidth]{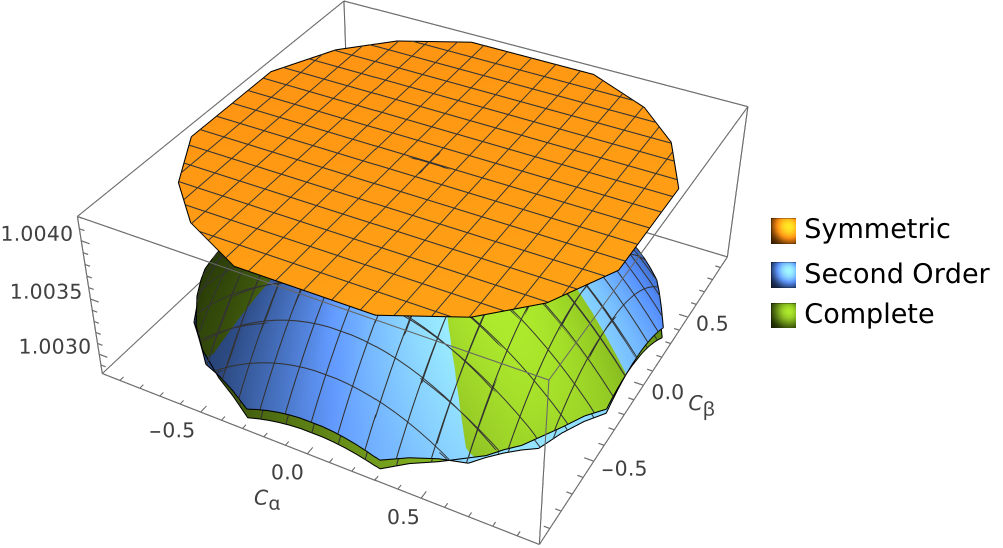}
    \caption{Graphical representation of the 4 dimensional form factor $\bar{\Gamma}_1$. The flat upper disc represents the symmetric approximation, while the overlapping surfaces are the second order expansion around the symmetric point and the complete expression. We fix $s^2=2.25$ GeV$^2$. The coordinates $Q^2$ and $R^2$ are normalized according to Eq. \eqref{eq:defs_ca_cb}. }
    \label{desarrollo4d}
\end{figure}
\section{Planar degeneracy of the three-gluon vertex\label{Sec:planar}}

The three-gluon vertex in the Landau gauge from the Curci-Ferrari model was computed both at one- and two-loop order \cite{Pelaez:2013cpa,Figueroa:2021sjm,Barrios:2022hzr}, by employing a purely perturbative framework within an infrared safe approach. These results are in very good agreement with the non-perturbative results from lattice simulations \cite{Parrinello:1994wd,Alles:1996ka,Cucchieri:2006tf,Cucchieri:2008qm,Athenodorou:2016oyh,Duarte:2016ieu,Boucaud:2017obn,Vujinovic:2018nqc,Pinto-Gomez:2022brg,Pinto-Gomez:2024mrk} as well as continuum approaches, such as those based on Dyson-Schwinger equations \cite{Eichmann:2014xya,Eichmann:2021zuv,Aguilar:2019jsj,Aguilar:2019uob,Aguilar:2021lke,Aguilar:2022thg,Aguilar:2023mdv,Aguilar:2023qqd,Huber:2014isa,Huber:2016tvc,Huber:2018ned,Huber:2020keu,Binosi:2017rwj} and the Functional Renormalization Group \cite{Cyrol:2016tym,Corell:2018yil}.

It is precisely the high level of accuracy of these perturbative calculations that allows us to provide an explanation for the planar degeneracy of the three-gluon vertex within the CF model.
At tree-level, planar degeneracy of the three-gluon vertex is fulfilled exactly, as it is entirely described by $\bar \Gamma_{1}=1$, whereas the remaining form factors vanish, see \eqref{generalbasis}. 

One-loop contributions could, in principle, introduce deviations from planar degeneracy larger than those observed in numerical simulations, owing to the moderate expansion parameter, which allows corrections up to one-third of the tree-level value. Nevertheless, as we will demonstrate in what follows, one-loop contributions respect planar degeneracy with remarkable precision. The reasons for this are, first, that the divergent logarithmic IR divergence is dominated by a single scale, thus automatically satisfying planar degeneracy and, second, that this divergence is only present in the tree-level form factor $\bar{\Gamma}_1$, which dominates over the remaining three form factors. Third, in the ultraviolet (UV) limit, asymptotic freedom ensures that the vertex recovers its classical structure, again dominated by a single scale. 
Finally, in the intermediate regime, on top of the moderate value of the perturbation theory expansion parameter, although the vertex is not perfectly planar, the Bose symmetry of the three-gluon vertex ensures that deviations at linear order in $Q^2$ and $R^2$ from the symmetric point (which trivially respects planar degeneracy) vanish. Consequently, the first non-trivial corrections to the symmetric point are of second order in these variables, making a quadratic expansion in the asymmetrical variables, $Q^2$ and $R^2$, sufficient to describe the one-loop vertex with high accuracy, as we will show in what follows.

\subsection{Planar degeneracy at one-loop order}

In this subsection, we will study the deviations of a general kinematic configuration from the symmetric point, which trivially fulfills planar degeneracy, for a given value of the kinematic variable $s^2$, at one-loop order. To this end, we first write the completely symmetric form factors, $\bar{\Gamma}_i(p^2,k^2,r^2)$, in terms of the variables $Q^2$, $R^2$ and $s^2$ as $\bar{\Gamma}_i(s^2,Q^2,R^2)$. Second, we expand each $\bar{\Gamma}_i$ in $Q^2$ and $R^2$ around the symmetric point, for a fixed $s^2$, which reads
\begin{align}\label{quadapprox}
\bar{\Gamma}_i&(s^2,Q^2,R^2)\approx f_0(s^2) + f_{1,Q^2}(s^2)Q^2 + f_{1,R^2}(s^2)R^2\nonumber\\
&+f_{2,Q^4}(s^2)Q^4+f_{2,R^4}(s^2)R^4+f_{2,Q^2 R^2}(s^2)Q^2 R^2+\cdots,
\end{align}
where $f_0(s^2)=\bar{\Gamma}_i(p^2,p^2,p^2)$ is the value of the form factor at the symmetric point $p^2=r^2=k^2$ and the ellipses denote higher order terms. Due to Bose symmetry, $f_{1,Q^2}(s^2)=f_{1,R^2}=f_{2,Q^2 R^2}(s^2)=0$ and $f_{2,R^4}(s^2)=3 f_{2,Q^4}(s^2)$.

The deviation from the symmetric approximation, which is a constant in the
$(Q^2,R^2)$ plane, provides an idea of how well planar degeneracy works as a
replacement for the full one-loop calculation. We define the mean relative
error to the symmetric approximation as
\begin{equation}
\sigma_{\mathrm{rel},i}(s^2)
  = \frac{\bigl|\bar\Gamma_i(p^2,p^2,p^2)-\langle\bar\Gamma_i\rangle\bigr|}
         {\bigl|\langle\bar\Gamma_i\rangle\bigr|},
         \label{eq:meanvalue}
\end{equation}
where 
\begin{equation}\label{eq:gamma_mean}
 \langle\bar\Gamma_i\rangle=\frac{1}{\Omega}\int_\Omega
\bar\Gamma_i(p^2,r^2,k^2)\,d\Omega,   
\end{equation}
with $i\in\{1,2,3,4\}$, is the mean value of the form factor for the allowed kinematics at fixed $s^2$.
On a surface of fixed $s^2$ the squared momenta are linear functions
of $Q^2$ and $R^2$, so the flat measure on $(p^2,r^2,k^2)$ restricted to that
surface is proportional to $dQ^2\,dR^2$, or equivalently to
$dC_\alpha\,dC_\beta$, where
\begin{equation}\label{eq:defs_ca_cb}
C_\alpha=\frac{\sqrt{3}Q^2}{s^2},\quad C_\beta=\frac{3R^2}{s^2}
\end{equation}
In these variables the allowed kinematic region is exactly the unit disk: $C_\alpha^2+C_\beta^2\leq 1$ \cite{Eichmann:2014xya,Aguilar:2023qqd}.

The integral is estimated by sampling the disk with $n$ points on a Vogel
spiral \cite{Vogel1979}. In polar coordinates,
$(C_\alpha, C_\beta) = \rho\,(\cos\theta, \sin\theta)$, with $\rho=0$ the
symmetric point and $\rho=1$ the collinear boundary, the points are
\begin{equation}
\rho_k = \rho_{\max}\sqrt{\frac{k-\frac12}{n}},\qquad
\theta_k = k\,\alpha,\qquad k=1,\dots,n,
\end{equation}
where $\alpha=\pi(3-\sqrt{5})$ is the golden angle. We opt for this sample because the radial law makes the points uniformly
distributed in area, so that $\langle\bar\Gamma_i\rangle$ is estimated by the mean over the sample.

The sample is kept away from the boundary, $\rho_{\max}<1$, since the transverse
basis degenerates there as well as on the three bisectoral lines where two of
the momenta coincide. In $d=3$, where all form factors are known in closed
form, we use $n=1200$ and $\rho_{\max}=0.999$, which leaves uncovered $0.2\%$ of
the area of the disk. In $d=4$ each evaluation requires a numerical
integration over Feynman parameters, and $\bar\Gamma_2$ results from a
cancellation between an analytic and an integrated contribution that becomes
severe near the boundary; there we use $n=200$ and $\rho_{\max}=0.95$ and
extrapolate, by using a quadratic fit in $1-\rho^2$ over nested disks, to the full disk. 

Applied in $d=3$, where the full-disk average is directly accessible,
the same procedure (fit restricted to $\rho\le 0.95$) reproduces
$\langle\bar\Gamma_1\rangle$ to within $2\times10^{-4}$ at all $s$, except in a narrow window around its zero crossing ($s\approx 0.5$~GeV), where
the relative difference is singular due to the vanishing denominator, see
$\sigma_{{\rm rel},1}$ in Eq.~\eqref{eq:meanvalue}. The corresponding uncertainty on
$\sigma_{\rm rel}$ is about $0.02$ percentage points for $\bar\Gamma_1$ and of
up to $1.7$ percentage points for $\bar\Gamma_{2,3,4}$.

In the case of $\bar \Gamma_1$, the error function \eqref{eq:meanvalue} would develop a singularity at the zero crossing, where $\langle \bar \Gamma_i\rangle \to 0$, thus losing the ability to measure the deviation from the symmetric configuration near this point. To circumvent this unphysical divergence, we adopt a combined error estimate tailored to the different momentum regimes:
\be\label{eq:combined_error}
\sigma_{\text{1}}(s^2) = 
\begin{cases}
\sigma_{\text{rel,1}}(s^2), & s^2 < s_{\text{zc}}^2, \\[8pt]
\dfrac{|\bar{\Gamma}_1(p^2,p^2,p^2) - \langle\bar{\Gamma}_1\rangle|}{|\bar{\Gamma}_0|}, & s^2 \ge s_{\text{zc}}^2,
\end{cases}
\ee
where $s_{\text{zc}}^2$ denotes a scale safely chosen below the zero-crossing momentum, and $\bar{\Gamma}_0$ is a characteristic reference value of the form factor evaluated at a chosen momentum scale away from zero. Both in three and four dimensions, we evaluate $\bar{\Gamma}_0$ at $s^2=1$ GeV$^2$. For the second-order approximation,
$\bar\Gamma_1^{(2)} = f_0(s^2) + f_{2,Q^4}(s^2)\,(Q^4 + 3R^4)$, the error is
instead averaged point by point,
\begin{equation}\label{eq:combined_error_expansion}
  \sigma_1^{(2)}(s^2) =
  \begin{cases}
    \left\langle \dfrac{|\bar\Gamma_1^{(2)} - \bar\Gamma_1|}{|\bar\Gamma_1|} \right\rangle, & s^2 < s_{\rm zc}^2,\\[3mm]
    \dfrac{\langle |\bar\Gamma_1^{(2)} - \bar\Gamma_1| \rangle}{|\bar\Gamma_0|}, & s^2 \ge s_{\rm zc}^2,
  \end{cases}
\end{equation}
since a difference of averages would miss deviations that cancel over the disk. In $d=3$, we derive the second order expansion analytically whereas in $d=4$ the coefficients $f_0$ and $f_{2,Q^4}$ of the second order expansion are extracted numerically: the
exact $\bar\Gamma_1$ is averaged over 12 equally spaced directions on circles of
radius $\rho = 0.15$, $0.2$, $0.25$ and $0.3$ around the symmetric point, which
removes the angular dependence, and the result is fitted to a polynomial in
$\rho^2$. Applied in $d=3$, this procedure reproduces the closed-form expansion
to better than $3\times10^{-5}$ over the whole disk.

In the deep infrared regime and in three dimensions ($s^2 < s_{\text{zc}}^2=0.14$ GeV$^2$), where the form factor $\bar{\Gamma}_1$ exhibits a divergence, the relative error $\sigma_{\text{rel}}(s^2)$ remains well behaved and provides a clear measure of the approximation accuracy\footnote{In this case, the absolute error defined in \eqref{eq:combined_error} for $s^2 \le s_{\text{zc}}^2$ is not a good measure of the discrepancy, since the numerator becomes much larger than the reference value $\bar{\Gamma}_0$ in the denominator.}. Around and above the zero-crossing scale, the normalized absolute error in \eqref{eq:combined_error} avoids artificial singularities by replacing the vanishing denominator with the constant energy scale reference $\bar{\Gamma}_0$. Although the relative error could in principle be used again at scales above the zero-crossing, both metrics yield nearly similar results because the form factor varies very slowly in that region. In four dimensions, since the infrared divergence of $\bar{\Gamma}_1$ is milder than in three dimensions, the absolute error is well behaved at all scales, consequently we choose $s_{\text{zc}}=0$.

\begin{figure*}[t]
  \centering
  \includegraphics[width=0.48\textwidth]{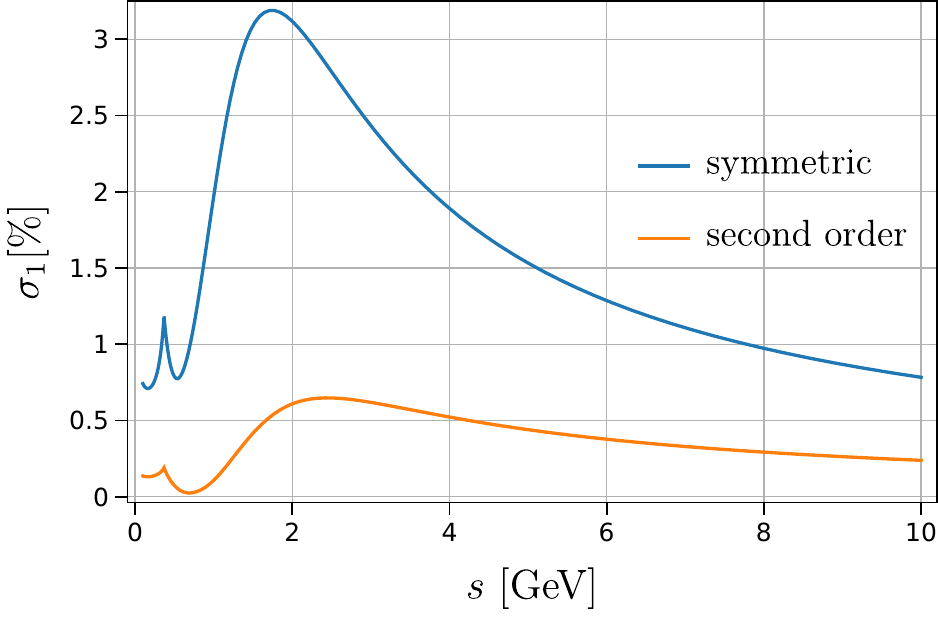}\hfill
  \includegraphics[width=0.48\textwidth]{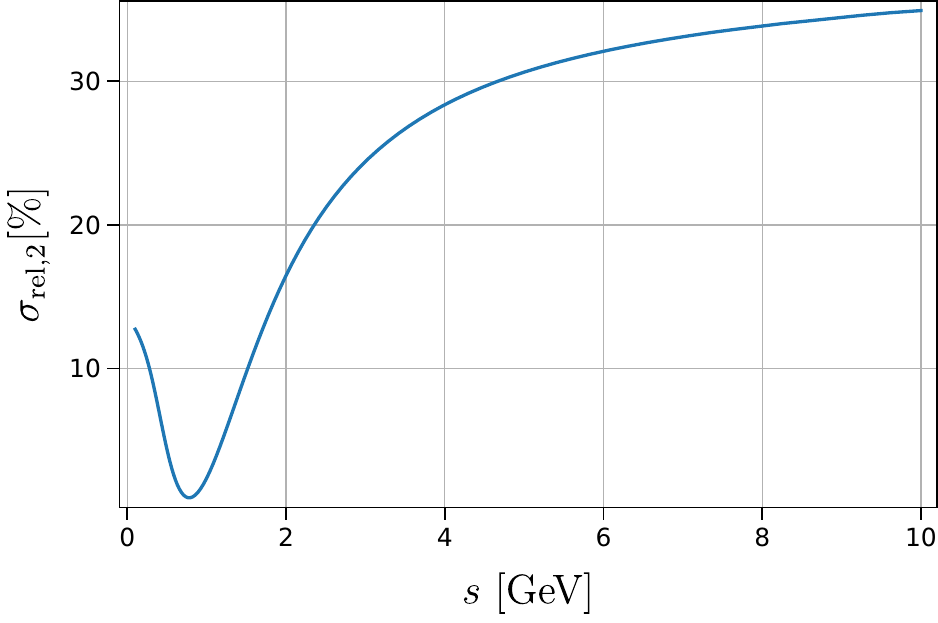}
  \includegraphics[width=0.48\textwidth]{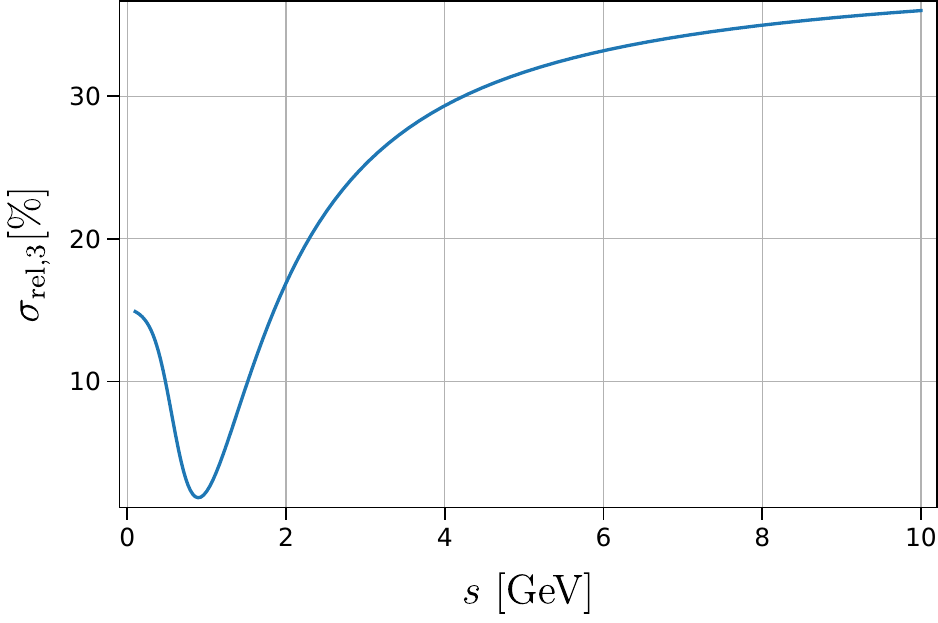}\hfill
  \includegraphics[width=0.48\textwidth]{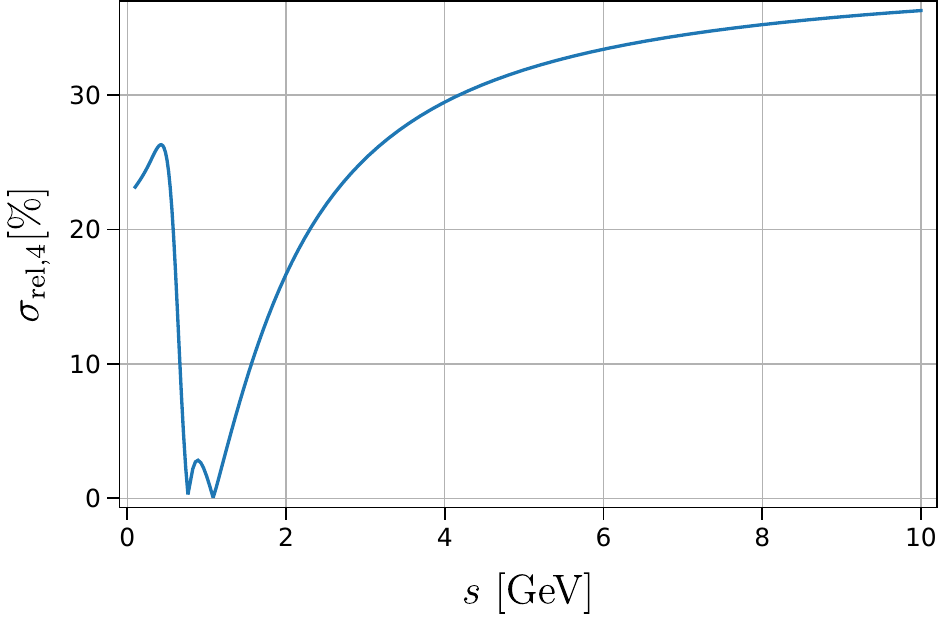}
  \caption{Relative error \eqref{eq:meanvalue} as a function of $s$ for the form factors $\bar \Gamma_2$ (top right), $\bar \Gamma_3$ (bottom left) and $\bar \Gamma_4$ (bottom right) in $d=3$. In the case of $\bar \Gamma_1$ (top left) the plot shows the error defined as \eqref{eq:combined_error} for the symmetric point and \eqref{eq:combined_error_expansion} for the second order expansion both with $s_{\text{zc}}^2=0.14$ GeV$^2$. The small spike at $s\approx 0.37$ GeV in the plot for $\bar \Gamma_1$ is where the two branches of the combined error \eqref{eq:combined_error} join.}
  \label{fig:sigmas_3d}
\end{figure*}

As we will discuss in the following subsections, for the tree-level tensor form factor $\bar{\Gamma}_1(p^2,r^2,k^2)$, the values attained do not deviate much from the symmetric point, see Figs. \ref{fig:sigmas_3d} and \ref{fig:sigmas_4d}. This corroborates the assertion given in \cite{Aguilar:2023qqd} that the tree-level form factor is essentially flat, i.e., independent of the coordinates $Q^2$ and $R^2$. This claim is valid for any value of $s^2$ both in three and four dimensions. When looking at the second order approximation, the deviation is much smaller, so that it can effectively replace the analytical expression with a high degree of accuracy. As for the other form factors $\bar{\Gamma}_i(p^2,r^2,k^2)$ ($i=2,3,4$), the deviation from the symmetric point is significant for larger values of $s^2$, especially around the soft-gluon configurations. This can be understood as a remnant of its UV behavior, where these points are plagued by collinear divergences. In the IR, the appearance of the gluon mass softens their behavior \cite{Pinto-Gomez:2022brg}. The larger deviations from the symmetric point are what remains of the now tamed divergences.

\begin{figure}
    \centering
    \includegraphics[width=\linewidth]{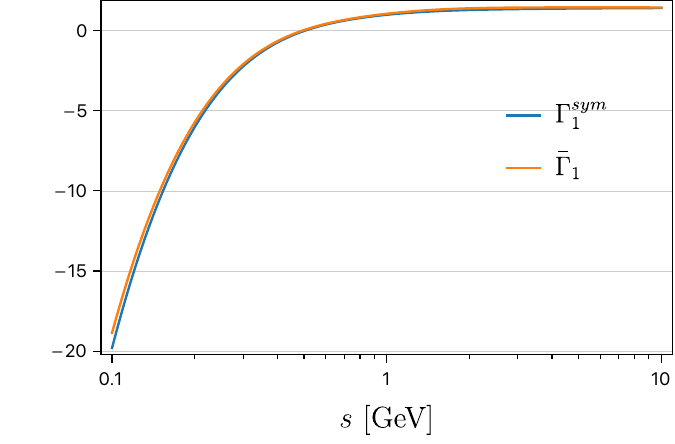}
    \caption{Comparison between the full symmetric form factor $\Gamma^{sym}_1$ and $\bar{\Gamma}_1$ in the symmetric limit for different values of momentum $s$ in three dimensions.}
    \label{gamma1sym}
\end{figure}

As a final comment, when considering the symmetric configuration form factor $\Gamma^{sym}_1$ \cite{Figueroa:2021sjm,Aguilar:2023qqd} proportional to the tree-level tensor, we find that $\Gamma^{sym}_1 \approx \bar{\Gamma}_1(q^2,q^2,q^2)$ is a reasonable approximation, given the small contributions of the other form factors and the fact that the IR divergence comes exclusively from $\bar{\Gamma}_1$. For a comparison between the full symmetric form factor and the $\bar{\Gamma}_1$ approximation in three dimensions, see Fig. \ref{gamma1sym}.

In the following subsections, we discuss the specific assumptions and results obtained by analyzing the three-gluon vertex in three and four dimensions. We do not focus on obtaining the values of the parameters $g_0$ and $m_0$ that best fit lattice data, as this has already been examined in previous works \cite{Pelaez:2013cpa,Figueroa:2021sjm,Barrios:2022hzr}, displaying very good agreement with simulations. Instead, our goal is to analyze the features associated with planar degeneracy, such as the flatness of the tree-level tensor form factor and the subleading behavior of the remaining form factors. In this respect, we not only corroborate previous findings from the lattice and DSE studies \cite{Pinto-Gomez:2022brg,Aguilar:2023qqd,Pinto-Gomez:2024mrk}, but provide a perturbative explanation for them, reinforcing the idea that some phenomena in the low energy regime of QCD/YM can be understood perturbatively when taking into account gluon mass effects \cite{Pelaez:2017bhh,Reinosa:2017qtf,Pelaez:2021tpq}. 

\subsection{Planar degeneracy in three dimensions}
\begin{figure*}
    \includegraphics[width=0.45\linewidth]{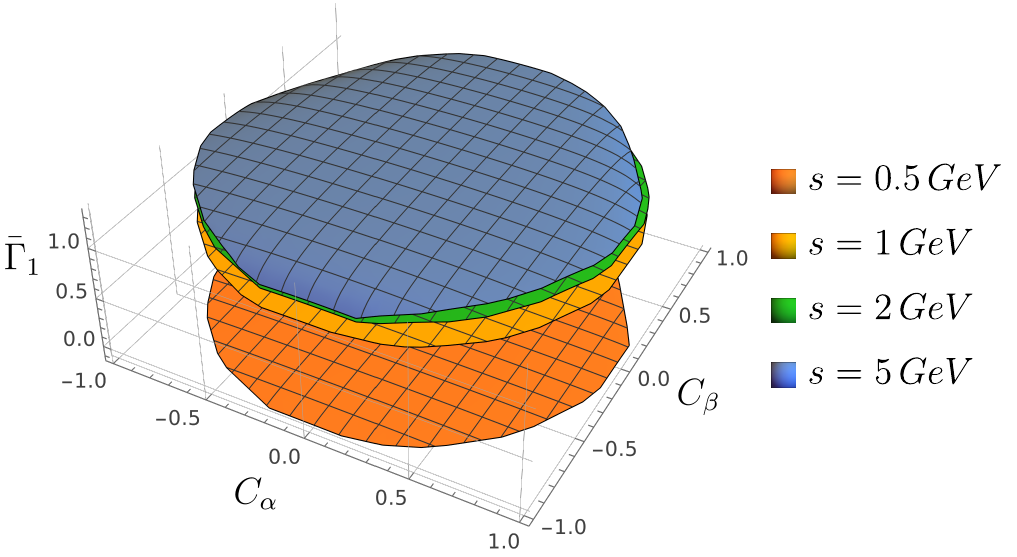}\hfill
    \includegraphics[width=0.45\linewidth]{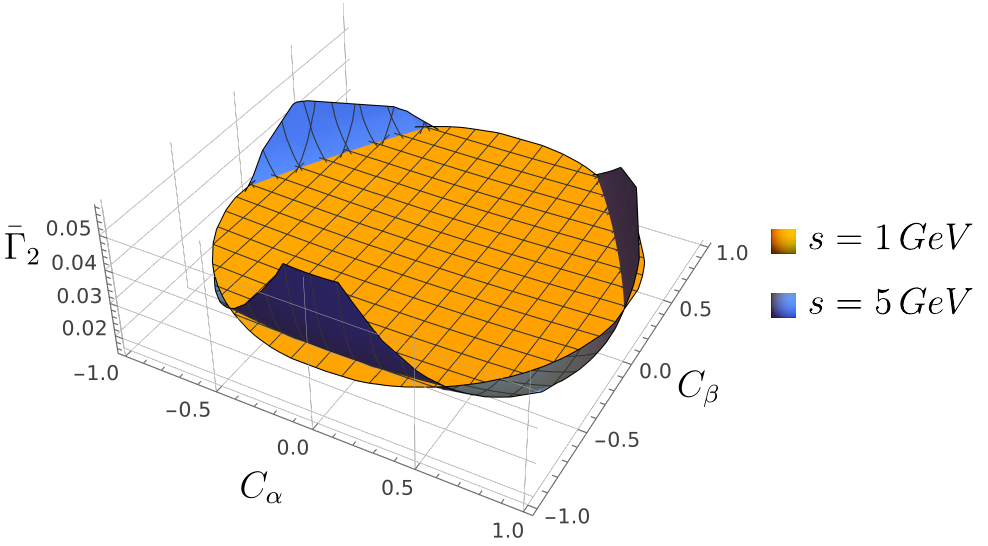}
    \includegraphics[width=0.45\linewidth]{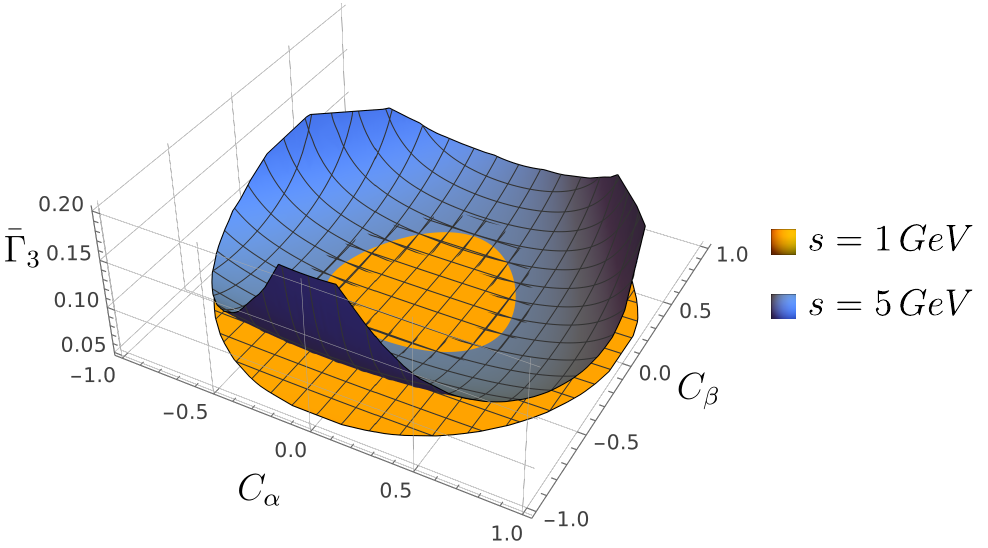}\hfill
    \includegraphics[width=0.45\linewidth]{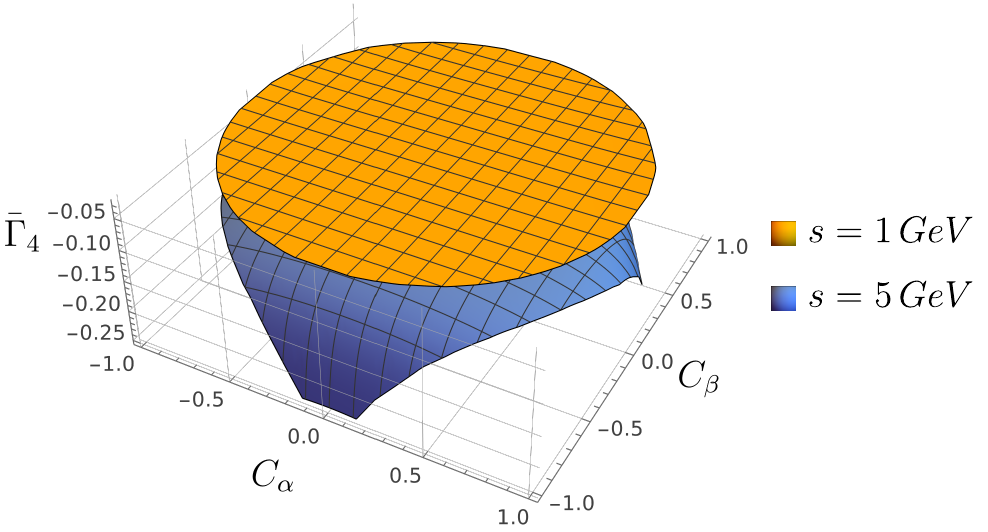}
    \caption{Form factors $\bar{\Gamma}_i(C_\alpha,C_\beta)$, with $i\in\{1,2,3,4\}$ at different values of $s$ in three dimensions.}
    \label{ff3d}
\end{figure*}

\begin{figure}[!t]
  \centering
  \includegraphics[width=0.48\textwidth]{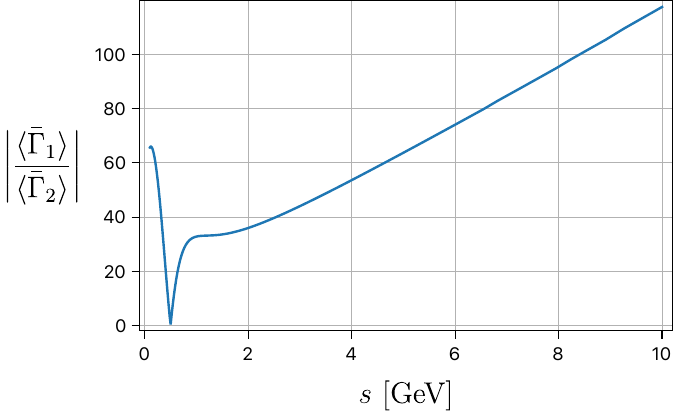}
  \includegraphics[width=0.48\textwidth]{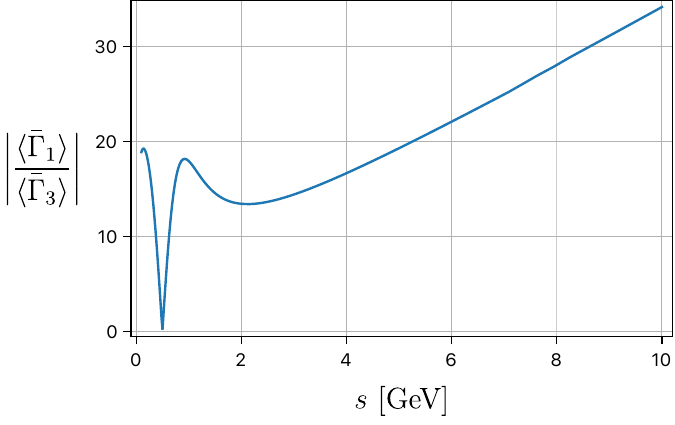}
  \includegraphics[width=0.48\textwidth]{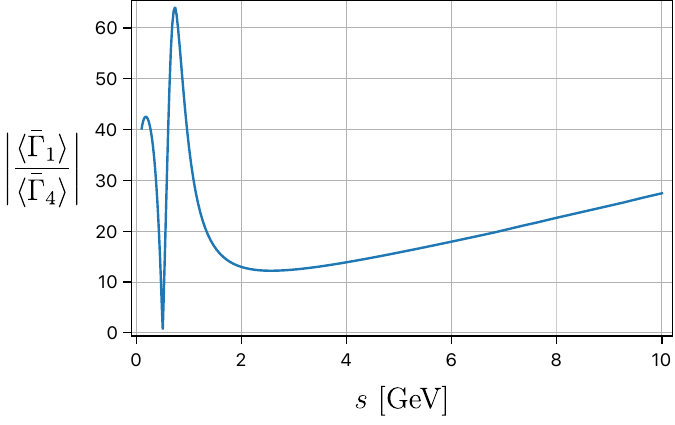}
  \caption{Ratio $|\langle \bar \Gamma_1\rangle$/$\langle \bar \Gamma_i\rangle|$, with $i\in\{2,3,4\}$ in three dimensions. The values $\langle\bar \Gamma_i\rangle$, including $\bar \Gamma_1$, are computed according to \eqref{eq:gamma_mean}.}
  \label{fig:comps_3d}
\end{figure}

To fix the initial parameters ($g_0,m_0$) in the infrared safe scheme, we choose the values that best fit the CF gluon and ghost propagators to the lattice data. For the three dimensional case, the initial parameters ($g_0,m_0$) chosen are $g_0=2.4$ GeV$^{1/2}$ and $m_0=0.55$ GeV \cite{Pelaez:2013cpa}, for $\mu_0=1$ GeV and the gauge group $SU(2)$. With these values, we evaluated each form factor $\bar \Gamma_{1,2,3,4}$, with the purpose of computing the deviation with respect to the symmetric configuration according to \eqref{eq:meanvalue} and \eqref{eq:combined_error}.

In Fig. \ref{fig:sigmas_3d}, we show the deviation with respect to the symmetric point for each form factor. As for $\bar \Gamma_1$, it grows up to $3.2\%$ around $s=1.8$ GeV, after which the running of the coupling constant forces the deviation  to be reduced, due to asymptotic freedom. This behavior is in line with the one observed in lattice simulations and DSE results \cite{Pinto-Gomez:2022brg,Aguilar:2023qqd}, which study the deviation from the soft-gluon configuration. In addition, the deviation from the second order expansion around the symmetric point grows up to $0.65\%$ around $s=2.4$ GeV, after which it also decreases towards the UV, though slower than in the symmetric case. The other form factors generally display large deviations with respect to the symmetric configuration, except for a short range of $s$, below $s=2$ GeV.

These conclusions are also supported by the plots of Fig. \ref{ff3d}, that display the form factors $\bar \Gamma_{1,2,3,4}$ as a function of the normalized variables $C_\alpha$ and $C_\beta$, defined in \eqref{eq:defs_ca_cb}. We observe that the form factor $\bar{\Gamma}_1$ remains almost flat over a wide range of values of $s$, while the other form factors display this behavior only for a range of values below $s=2$ GeV. For values $s\gtrsim 2$ GeV, $\bar \Gamma_{2,3,4}$ show significant deviations from the symmetric point, especially around the soft-gluon configurations. 

The results of Figs. \ref{fig:sigmas_3d} and \ref{ff3d} show that at one-loop order, planar degeneracy is an excellent approximation for $\bar \Gamma_1$, but it performs poorly in the case of the remaining form factors. As a consequence, to test the property of planar degeneracy on the three gluon vertex as a whole, we must study the relative weight of $\bar \Gamma_{2,3,4}$ in relation to $\bar \Gamma_1$. To that end, we computed $|\langle \bar \Gamma_1\rangle/\langle \bar \Gamma_i \rangle|$, with $i\in \{2,3,4\}$ as a function of $s$. The results are displayed in Fig. \ref{fig:comps_3d}, which, with the exception of the region close to the zero crossing of $\bar \Gamma_1$, show a clear sub-dominance of $\bar{\Gamma}_{2,3,4}$ with respect to $\bar{\Gamma}_1$. Moreover, from Figs. \ref{fig:sigmas_3d} and \ref{fig:comps_3d}, away from the zero crossing of $\bar \Gamma_1$, $\sigma_{\text{rel},i}\times |\langle \bar \Gamma_i\rangle/\langle \bar \Gamma_1\rangle|$ is at most about $2\%$ for $i\in\{2,3,4\}$, so the non-planarity that $\bar \Gamma_{2,3,4}$ add to the full vertex is of the same order as that of $\bar \Gamma_1$ itself.

Additionally, considering that the largest deviations from the symmetric point for $\bar \Gamma_{2,3,4}$ occur for soft gluon configurations and that, as argued in \cite{Aguilar:2023qqd}, the basis elements $\tau^i_{\mu\nu\rho}, (i=2,3,4)$ vanish in the soft gluon limit linearly with the vanishing momentum, the approximation

\be\label{planardeg}
\Bar{\Gamma}_{\mu\nu\rho}(p,r,k) = \bar{\Gamma}^{1}(2 s^2/3,2 s^2/3,2s^2/3)\bar{\Gamma}_{\mu\nu\rho}^0
\ee
is reasonable even in these limits.

Since planar degeneracy provides an accurate description of the vertex when approximated by the form factor in the symmetric configuration, we present the corresponding analytical expressions for the symmetric case in Appendix~\ref{App:expressions}. Furthermore, the second-order corrections and the explicit expressions for the vertex in arbitrary kinematics are made available in the Supplemental Material \cite{supplementalMaterial}.


\subsection{Planar degeneracy in four dimensions}

\begin{figure*}[t]
  \centering
  \includegraphics[width=0.48\textwidth]{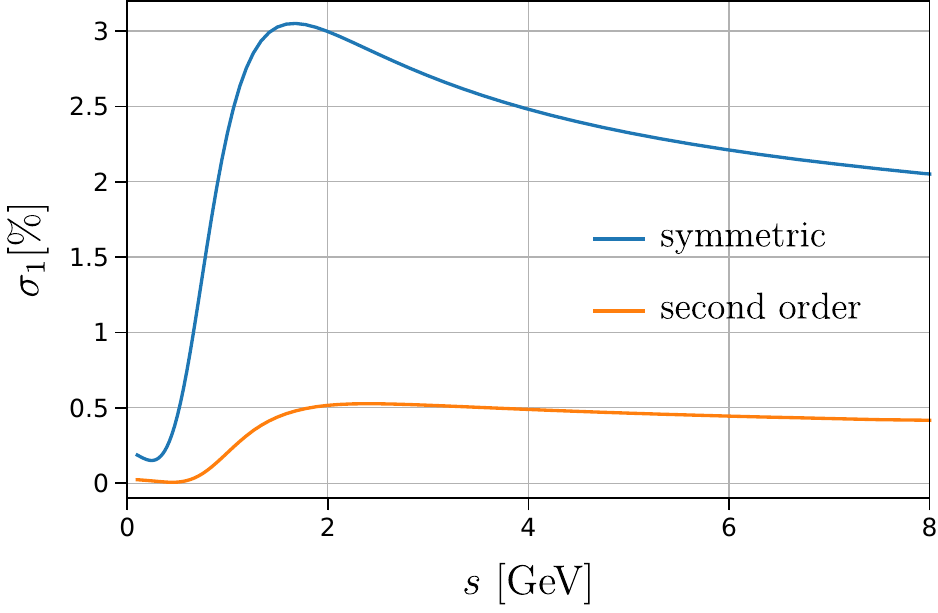}\hfill
  \includegraphics[width=0.48\textwidth]{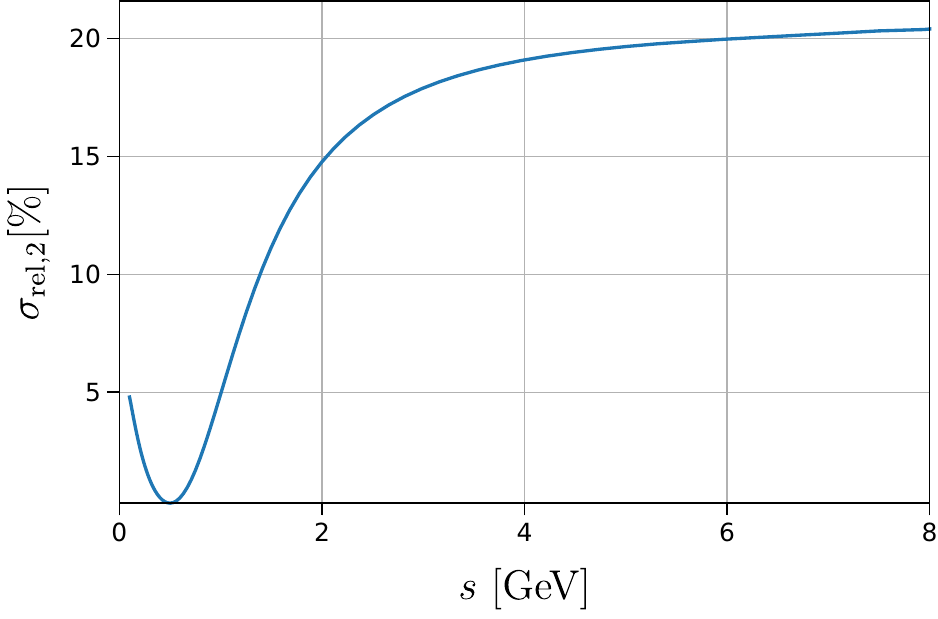}
  \includegraphics[width=0.48\textwidth]{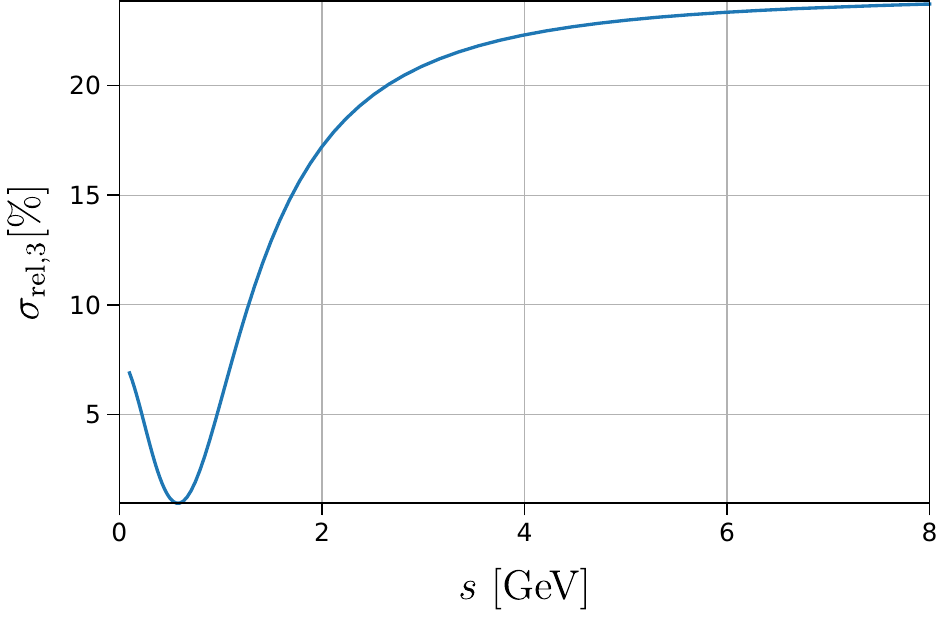}\hfill
  \includegraphics[width=0.48\textwidth]{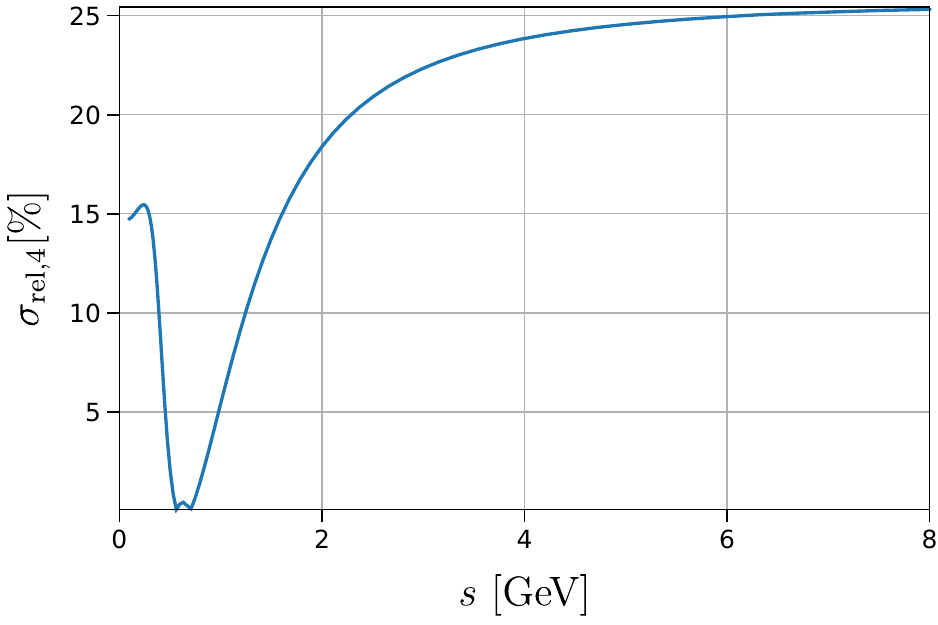}
  \caption{Relative error \eqref{eq:meanvalue} as a function of $s$ for the form factors $\bar \Gamma_2$ (top right), $\bar \Gamma_3$ (bottom left) and $\bar \Gamma_4$ (bottom right) in d=4. In the case of $\bar \Gamma_1$ (top left) the plot shows the deviation \eqref{eq:combined_error} for the symmetric point and \eqref{eq:combined_error_expansion} for the second order expansion, both with $s_{\text{zc}}=0$.}
  \label{fig:sigmas_4d}
\end{figure*}

\begin{figure*}[t]
    \includegraphics[width=0.45\linewidth]{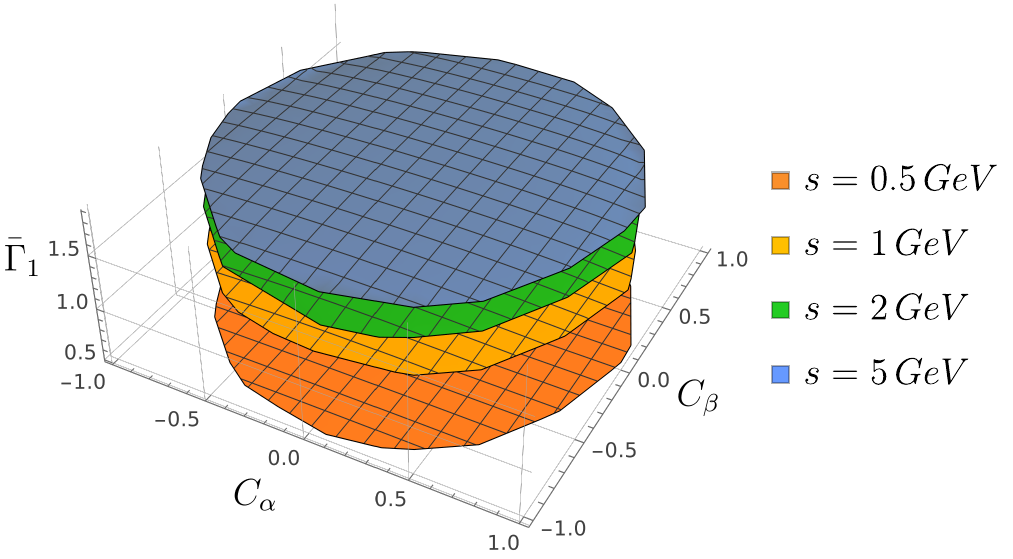}\hfill
    \includegraphics[width=0.45\linewidth]{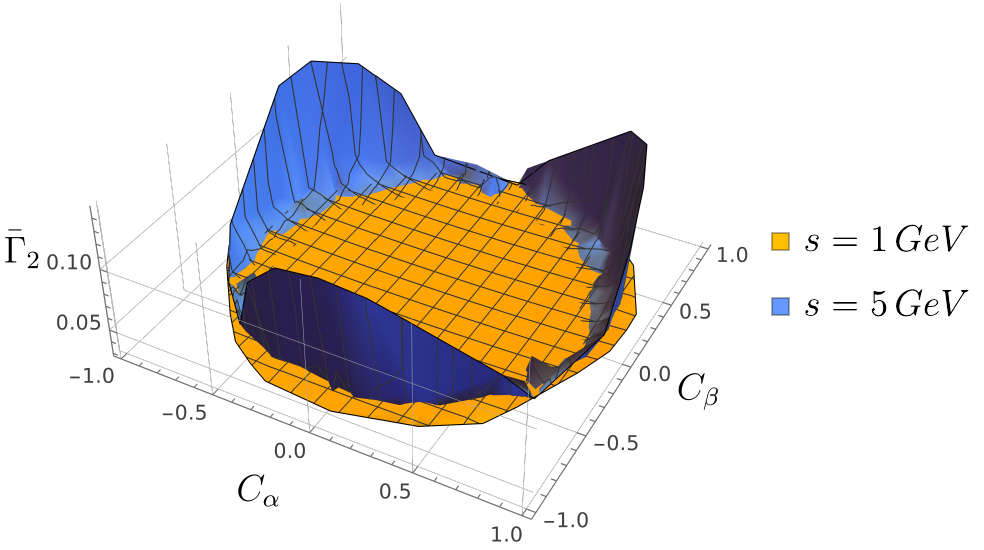}
    \includegraphics[width=0.45\linewidth]{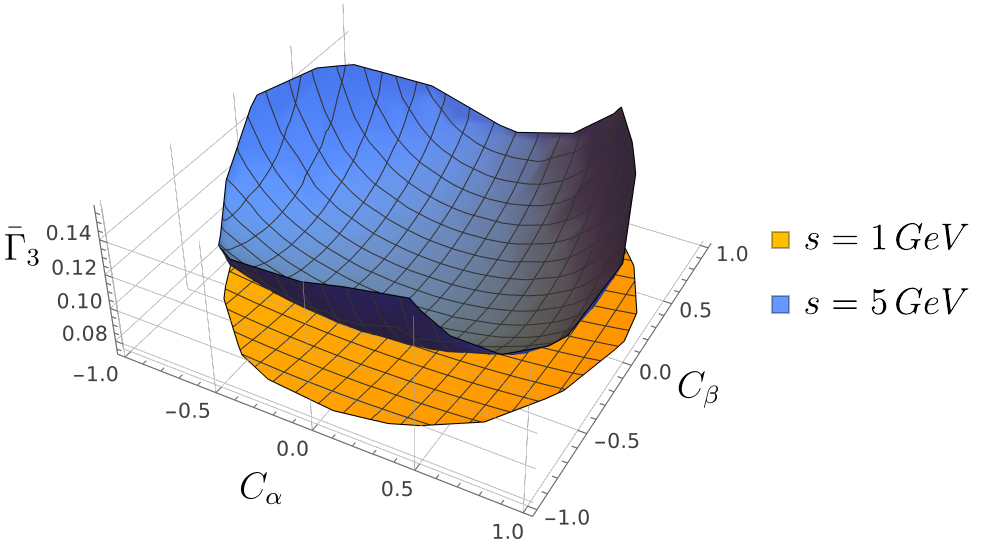}\hfill
    \includegraphics[width=0.45\linewidth]{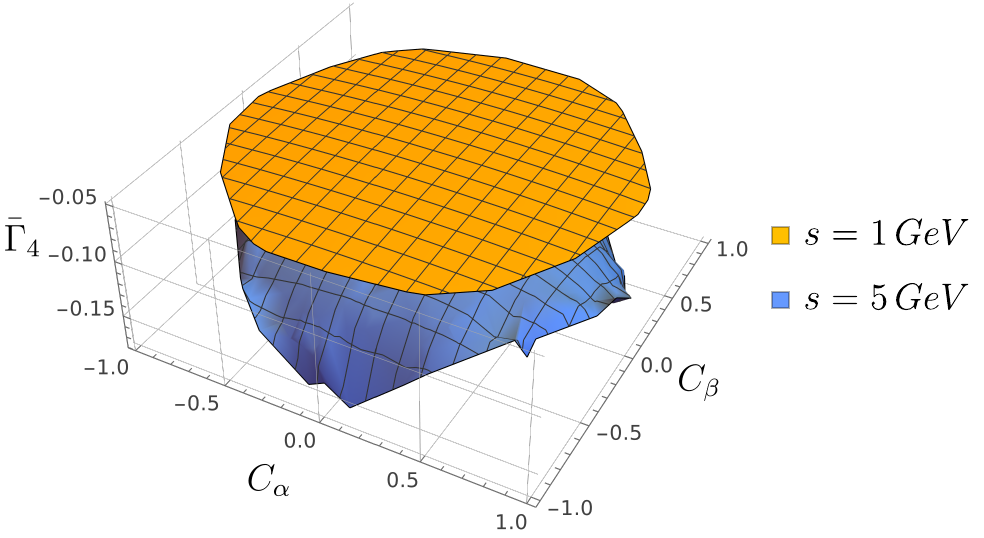}
    \caption{Form factors $\bar{\Gamma}_i(C_\alpha,C_\beta)$, with $i\in\{1,2,3,4\}$, at different values of $s$ in four dimensions.}
    \label{ff4d}
\end{figure*}

\begin{figure}[!t]
  \centering
  \includegraphics[width=0.48\textwidth]{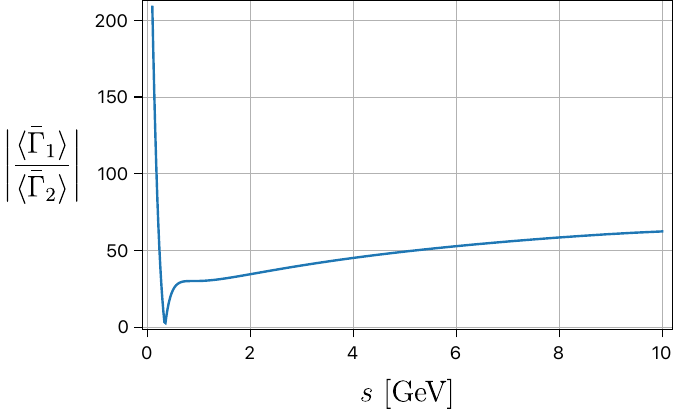}
  \includegraphics[width=0.48\textwidth]{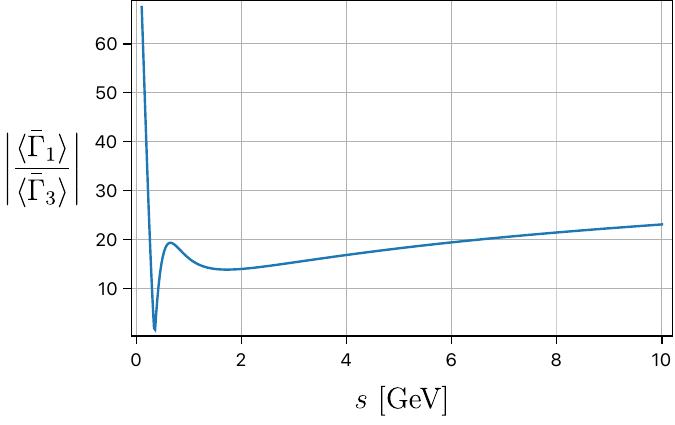}
  \includegraphics[width=0.48\textwidth]{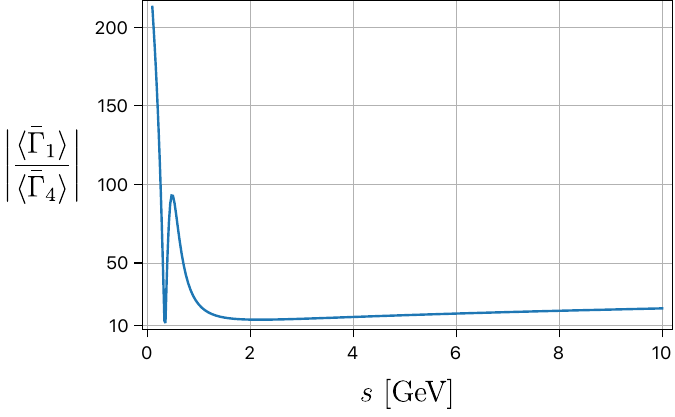}
  \caption{Ratio $|\langle \bar \Gamma_1\rangle$/$\langle \bar \Gamma_i\rangle|$, with $i\in\{2,3,4\}$ in four dimensions. The values $\langle\bar \Gamma_i\rangle$, including $\bar \Gamma_1$, are computed according to \eqref{eq:gamma_mean}.}
  \label{fig:comps_4d}
\end{figure}

In four dimensions, for a fixed renormalization scale $\mu_0=1$ GeV and the gauge group $SU(3)$, the best fit with the lattice data is given by the initial parameters $g_0=3.6$ and $m_0=0.35$ GeV \cite{Gracey:2019xom,Figueroa:2021sjm}. In what follows, we perform the same analysis as in the three dimensional case.

The deviations from the symmetric point, displayed in Fig.~\ref{fig:sigmas_4d}, feature the same qualitative behavior as in three dimensions. As for $\bar \Gamma_1$, the deviation rises sharply from the IR, dropping slowly towards the UV, but milder than in three dimensions, with a peak of $3\%$ around $s=1.8$ GeV. In the case of the deviation with respect to the second order approximation, it just surpasses the value of $0.5\%$, corroborating that, just as in three dimensions, it agrees with the exact one-loop result with high accuracy. 

In the case of $\bar \Gamma_{2,3,4}$, we observe that the value of $\sigma$ diminishes towards the region of the zero crossing of $\bar \Gamma_1$, reaching a minimum, and increasing to a value of $20-25\%$ towards the UV. 

Using the same normalized variables defined in \eqref{eq:defs_ca_cb}, we show the form factors $\bar{\Gamma}_{1,2,3,4}$ in the plane determined by $C_\alpha$ and $C_\beta$ for various values of $s$ in Fig. \ref{ff4d}. The picture is the same as in three dimensions. The form factor $\bar \Gamma_1$ features a flat behavior for all values of $s$ whereas $\bar \Gamma_{2,3,4}$ present this behavior just in the case of $s=1$ GeV, in agreement with the plots of Fig.~\ref{fig:sigmas_4d}. 

The ratios between $|\langle \bar \Gamma_1\rangle|$ and $|\langle \bar \Gamma_i\rangle|$ with $i\in \{2,3,4\}$ as a function of $s$ are provided in Fig.~\ref{fig:comps_4d}. Similarly to the three dimensional case $\bar\Gamma_1$ is clearly dominant over the remaining three form factors, with the exception of the region close to the zero crossing. It is also clear from Figs. \ref{fig:sigmas_4d} and \ref{fig:comps_4d} that $\sigma_{\text{rel},i}\times |\langle \bar \Gamma_i\rangle/\langle \bar \Gamma_1\rangle|$ for $i\in\{2,3,4\}$ is at most about $2\%$ (away from the zero crossing of $\bar \Gamma_1$), around the same deviation of $\bar \Gamma_1$ itself. Thus, we can conclude that planar degeneracy is an approximate property of the complete three-gluon vertex in four dimensions as well.

For completeness, we provide the symmetric expression of the form factor $\bar{\Gamma}_1$ in the Appendix \ref{App:expressions}, whereas the complete expression is given in the supplemental material \cite{supplementalMaterial}.

\section{Conclusions}\label{Sec:conclusions}
In this work, we have analyzed the planar degeneracy of the transversely projected three-gluon vertex within the framework of the Curci-Ferrari model. Our investigation shows that the infrared dynamics of the vertex are dominated by the tree-level tensor structure, which encapsulates the infrared divergence and remains essentially independent of the asymmetric kinematical variables. By expanding the form factors around the symmetric point, we found that Bose symmetry forces the first non-trivial corrections to be of second order, which explains the high degree of flatness observed in the vertex surfaces. While minor deviations appear in the intermediate regime where the coupling constant reaches its maximum, the running of the coupling towards the ultraviolet ensures that the vertex recovers its classical structure through asymptotic freedom. We conclude that the planar degeneracy is a robust feature of Yang-Mills theory in both three and four dimensions that can be understood perturbatively once a gluon mass scale is present. These findings further validate the Curci-Ferrari model as a powerful analytical tool that remains computationally tractable while providing results in excellent agreement with non-perturbative benchmarks.

\section*{Acknowledgments}
We thank Urko Reinosa and Matthieu Tissier for very useful discussions related to this work. The authors acknowledge financial support from the PEDECIBA program, from the ANII-FCE-2025-186497 project and the Ministry of Education and Culture of Uruguay through the project FVF 2023 513. 
\appendix

\section{Analytical expressions}
\label{App:expressions}
In this appendix, we list the analytical expressions for the 1-loop three gluon vertex in the symmetrical configuration. Although these expressions exist for general momentum configurations, both in three and four dimensions, they are rather long and cumbersome, so we collect them in an external file, which can be found in the supplementary materials \cite{supplementalMaterial}. Furthermore, our comparisons throughout the text with the symmetric limit motivate the introduction of the expressions here.

In three dimensions, the $\bar{\Gamma}_1(p^2)$ symmetric expression reads

\begin{widetext}
\begin{equation}\label{gamma1symd3}
\begin{aligned}
   \bar{\Gamma}_1(\tilde p^2)= 1+\frac{\tilde g^2 N }{24576 \pi }\left(c_1 \cot ^{-1}\left(\frac{3 \tilde{p}^2+8}{\tilde{p}^2 \sqrt{\tilde{p}^2+3}}\right)+c_2 \cot
   ^{-1}\left(\frac{\tilde{p}}{\sqrt{\tilde{p}^4+\tilde{p}^2+1}}\right)+c_3 \tan ^{-1}\left(\frac{\tilde{p}}{2}\right)+c_4 \tan
   ^{-1}(\tilde{p})+t_1+t_2\right)
\end{aligned}
\end{equation}
with
\begin{align}
    t_1&=-\frac{\pi  \left(102 \tilde{p}^6+1575 \tilde{p}^4+718 \tilde{p}^2-408\right)}{\tilde{p}^3}\,\nonumber\\
    t_2&=-\frac{12 \left(329 \tilde{p}^{14}+1958 \tilde{p}^{12}+3013 \tilde{p}^{10}+3606 \tilde{p}^8+1919 \tilde{p}^6+1548 \tilde{p}^4+566 \tilde{p}^2+600\right)}{\tilde{p}^4
   \left(\tilde{p}^2+4\right) \left(\tilde{p}^4+\tilde{p}^2+1\right)^2}\,\nonumber\\
   c_1&=\frac{256 \left(27 \tilde{p}^2+64\right)}{\sqrt{\tilde{p}^2+3}}\,\nonumber\\
   c_2&=\frac{16 \left(108 \tilde{p}^{14}+484 \tilde{p}^{12}+943 \tilde{p}^{10}+1045 \tilde{p}^8+662 \tilde{p}^6+170 \tilde{p}^4-49 \tilde{p}^2-51\right)}{\tilde{p}^3
   \left(\tilde{p}^4+\tilde{p}^2+1\right)^{5/2}}\,\nonumber\\
   c_3&=-\frac{6 \left(34 \tilde{p}^8+3555 \tilde{p}^6+3854 \tilde{p}^4+304 \tilde{p}^2+160\right)}{\tilde{p}^5}\,\nonumber\\
   c_4&=\frac{8 \left(51 \tilde{p}^8+1980 \tilde{p}^6+1564 \tilde{p}^4-192 \tilde{p}^2+285\right)}{\tilde{p}^5}
\end{align}
\end{widetext}
 Notice that the three gluon vertex is UV finite in $d=3$, therefore no pole appears in expression \eqref{gamma1symd3}. While the infrared behavior reads:
 \begin{align}
     \bar{\Gamma}_1(\tilde p^2)\sim -\frac{15 \left(\tilde{g}^2 N\right)}{4096 \tilde{p}}+\left(1-\frac{7 \tilde{g}^2 N}{48 \pi }\right)+O\left(\tilde{p}^2\right)
 \end{align}
where $\tilde p=\frac{p}{m}$ and $\tilde g^2=\frac{g^2}{m}$.
For completeness, we also give the vanishing momentum expression of the vertex correcting Eq.(19) of \cite{Pelaez:2013cpa}, where a typo was found.

\begin{widetext}
\begin{equation}
\label{eq_3gluonexceptional_3d}
\begin{split}
  \Gamma_{\mu\nu\rho}(p,0,-p)=&\bigg\{1 -\frac{N g_0^2}{128\pi m\sqrt s}\bigg[
  \frac\pi 2 (2-3s^2)+\frac{2}{3s^{3/2}}(3s^3+23s^2+56s-15)\\&
  -s^{-1}(4+s)(16-18s+s^2)\arctan(\sqrt s/2)
  +2s^{-2}(s-5)(s+1)^2(2s-1)\arctan(\sqrt s)\bigg]\bigg\} (p_\mu\delta_{\nu\rho}+p_\rho\delta_{\mu\nu}) \\
  & - \bigg\{ 2 +\frac{N g_0^2}{128\pi m\sqrt s}\bigg[
  \pi (-s^3+3s^2-1)+\frac{2}{3s^{3/2}}(15s^3-51s^2+53s-15)\\&+
  2(-s^3+6s^2+2s-16-32 s^{-1})\arctan(\sqrt s/2)
  +2(2 s^3 - 9 s^2 - 5)(1-s^{-2})\arctan(\sqrt s)\bigg]\bigg\} p_\nu\delta_{\mu\rho}\\
& -\frac{N g_0^2}{128\pi m^3 \sqrt s}\bigg[
  \pi  (s^{-1}+s^2)+\frac{2}{3s^{5/2}}(-21s^3+5s^2-139s+75)\\& +2 s^{-2}(s^4-5s^3-16s^2-40s+96)\arctan(\sqrt s/2) \\&
  -2s^{-3}(s+1)(2s^4-7s^3-9s^2-15s+25)\arctan(\sqrt s)\bigg] p_\mu p_\nu p_\rho
\end{split}
\end{equation}
\end{widetext}
In this case $s=p^2/m^2$, so we keep the notation from \cite{Pelaez:2013cpa}.
The analytical expression in other kinematical configurations is given in the supplemental material \cite{supplementalMaterial}.

In four dimensions, the $\bar{\Gamma}_1(p^2)$ symmetric expression, before renormalization, reads
\begin{widetext}
\begin{equation}\label{gamma1symd4}
\begin{aligned}
\bar{\Gamma}_1(p^2)&= 1 + g^2 N \Biggl(
-\frac{17}{96\,\pi^2\,\varepsilon}
+\frac{17\gamma_E}{192\,\pi^2}
-\frac{(-108m^6+234m^4p^2+685m^2p^4+160p^6)}{2304\,p^4(4m^2+p^2)\,\pi^2}
\\
&+\frac{\left(24m^{12}+200m^{10}p^2+2070m^8p^4+2706m^6p^6+887m^4p^8+60m^2p^{10}-6p^{12}\right)\operatorname{arctanh}\!\left(\frac{\sqrt{p^2}}{\sqrt{4m^2+p^2}}\right)}{384\,m^4p^4\sqrt{p^2}\,(4m^2+p^2)^{3/2}\,\pi^2}
\\
&-\frac{\left(8m^8+45m^6p^2+132m^4p^4-379m^2p^6-123p^8\right)\operatorname{Cl}_2\!\left(\frac{\pi}{3}\right)}{576\sqrt{3}\,m^2p^6\,\pi^2}
\\
&-\frac{(8m^2+3p^2)\left(\,\operatorname{Cl}_2\!\left(\frac{\pi}{3}-\arccos\frac{6m^2+p^2}{6m^2+2p^2}\right)+\,\operatorname{Cl}_2\!\left(\frac{\pi}{3}+\arccos\frac{6m^2+p^2}{6m^2+2p^2}\right)\right)}{16\sqrt{3}\,m^2\pi^2}
\\
&-\frac{(m^2+p^2)^2(2m^4+5m^2p^2+21p^4)\left(\,\operatorname{Cl}_2\!\left(\frac{\pi}{3}-2\arctan\frac{\sqrt{1+4m^2/p^2}}{\sqrt{3}}\right)+\,\operatorname{Cl}_2\!\left(\frac{\pi}{3}+2\arctan\frac{\sqrt{1+4m^2/p^2}}{\sqrt{3}}\right)\right)}{192\sqrt{3}\,m^2p^6\,\pi^2}
\\
&-\frac{(m^6+m^2p^4+p^6)\left(\,\operatorname{Cl}_2\!\left(\frac{\pi}{3}-2\arctan\frac{2m^2+p^2}{\sqrt{3}\,p^2}\right)+2\,\operatorname{Cl}_2\!\left(\frac{\pi}{3}+2\arctan\frac{2m^2+p^2}{\sqrt{3}\,p^2}\right)+\,\operatorname{Cl}_2\!\left(\pi-2\arctan\frac{2m^2+p^2}{\sqrt{3}\,p^2}\right)\right)}{128\sqrt{3}\,m^2p^4\,\pi^2}
\\
&-\frac{1}{768\,m^4p^6\pi^2}\Bigl[
  \left(6m^6p^4+m^4p^6+6m^2p^8\right)\ln\frac{p^2}{m^2}+6p^{10}\ln\frac{p^2}{m^2+p^2}
\\
  &-2\left(4m^{10}-6m^8p^2+60m^6p^4+121m^4p^6+45m^2p^8-3p^{10}\right)\ln\frac{m^2}{m^2+p^2}
  +68\,m^4p^6\ln\frac{4\pi}{m^2}
\Bigr]
\Biggr),
\end{aligned}
\end{equation}


\end{widetext}
where $\operatorname{Cl}_2(x)$ is the Clausen function of order 2. The factor $\ln\frac{4\pi}{m^2}$ appearing in the last term of the expression \eqref{gamma1symd4} is due to the dimensional regularization procedure, where the inclusion of a $\mu^\varepsilon$ factor would amount to a rewriting of the logarithm as $\ln\frac{4\pi \mu^2}{m^2}$.

Even though there are analytical expressions for the general configuration form factors, in both three and four dimensions, they are too long to be put here. Moreover, the 4 dimensional case is expressed in terms of generalized hypergeometric functions \cite{Davydychev:1990cq,Davydychev:1990jt}, which are not particularly elucidative nor easy to deal with numerically. For these reasons, we refrain to write them explicitly.

\bibliography{refs}

@article{Vogel1979,
  author  = {Vogel, Helmut},
  title   = {A better way to construct the sunflower head},
  journal = {Mathematical Biosciences},
  volume  = {44},
  number  = {3-4},
  pages   = {179--189},
  year    = {1979},
  doi     = {10.1016/0025-5564(79)90080-4}
}

@article{Pelaez:2022rwx,
    author = "Pel{\'a}ez, Marcela and Reinosa, Urko and Serreau, Julien and Wschebor, Nicol{\'a}s",
    title = "{Small parameters in infrared QCD: The pion decay constant}",
    eprint = "2212.10089",
    archivePrefix = "arXiv",
    primaryClass = "hep-ph",
    doi = "10.1103/PhysRevD.107.054025",
    journal = "Phys. Rev. D",
    volume = "107",
    number = "5",
    pages = "054025",
    year = "2023"
}

@book{Weinberg:1995mt,
    author = "Weinberg, Steven",
    title = "{The Quantum theory of fields. Vol. 1: Foundations}",
    doi = "10.1017/CBO9781139644167",
    isbn = "978-0-521-67053-1, 978-0-511-25204-4",
    publisher = "Cambridge University Press",
    month = "6",
    year = "2005"
}

@article{Pelaez:2020ups,
    author = "Pel{\'a}ez, Marcela and Reinosa, Urko and Serreau, Julien and Tissier, Matthieu and Wschebor, Nicol{\'a}s",
    title = "{Spontaneous chiral symmetry breaking in the massive Landau gauge: realistic running coupling}",
    eprint = "2010.13689",
    archivePrefix = "arXiv",
    primaryClass = "hep-ph",
    doi = "10.1103/PhysRevD.103.094035",
    journal = "Phys. Rev. D",
    volume = "103",
    number = "9",
    pages = "094035",
    year = "2021"
}

@article{Oribe:2025ezp,
    author = "Oribe, Santiago and Pel{\'a}ez, Marcela and Reinosa, Urko",
    title = "{Infrared-safe Minkowskian Curci-Ferrari model}",
    eprint = "2503.14291",
    archivePrefix = "arXiv",
    primaryClass = "hep-ph",
    doi = "10.1103/gxg3-5vxr",
    journal = "Phys. Rev. D",
    volume = "112",
    number = "1",
    pages = "014005",
    year = "2025"
}

@misc{Alvez:2025wek,
    author = "Alvez, Anaclara and Barrios, Nahuel and Ben{\'\i}tez, Florencia and Pel{\'a}ez, Marcela",
    title = "{Nonrelativistic meson masses from the Curci-Ferrari model}",
    eprint = "2509.04365",
    archivePrefix = "arXiv",
    primaryClass = "hep-ph",
    month = "9",
    year = "2025"
}

@article{Maelger:2018vow,
    author = "Maelger, Jan and Reinosa, Urko and Serreau, Julien",
    title = "{Universal aspects of the phase diagram of QCD with heavy quarks}",
    eprint = "1805.10015",
    archivePrefix = "arXiv",
    primaryClass = "hep-th",
    doi = "10.1103/PhysRevD.98.094020",
    journal = "Phys. Rev. D",
    volume = "98",
    number = "9",
    pages = "094020",
    year = "2018"
}

@article{Maelger:2019cbk,
    author = "Maelger, J. and Reinosa, U. and Serreau, J.",
    title = "{Localized rainbows in the QCD phase diagram}",
    eprint = "1903.04184",
    archivePrefix = "arXiv",
    primaryClass = "hep-th",
    doi = "10.1103/PhysRevD.101.014028",
    journal = "Phys. Rev. D",
    volume = "101",
    number = "1",
    pages = "014028",
    year = "2020"
}

@article{Surkau:2024zfb,
    author = "Surkau, V. Tomas Mari and Reinosa, Urko",
    title = "{Deconfinement transition within the Curci-Ferrari model: Renormalization scale and scheme dependences}",
    eprint = "2401.17869",
    archivePrefix = "arXiv",
    primaryClass = "hep-ph",
    doi = "10.1103/PhysRevD.109.094033",
    journal = "Phys. Rev. D",
    volume = "109",
    number = "9",
    pages = "094033",
    year = "2024"
}

@article{Pelaez:2024mtq,
    author = "Pel{\'a}ez, Marcela and Reinosa, Urko and Serreau, Julien and Tissier, Matthieu and Wschebor, Nicol{\'a}s",
    title = "{Scenario for quark confinement from infrared safe Yang-Mills dynamics}",
    eprint = "2405.20532",
    archivePrefix = "arXiv",
    primaryClass = "hep-th",
    doi = "10.1103/rk6n-t66l",
    journal = "Phys. Rev. D",
    volume = "112",
    number = "9",
    pages = "094017",
    year = "2025"
}

@misc{MariSurkau:2026irs,
    author = "Mari Surkau, V. Tomas and Reinosa, Urko",
    title = "{Interplay between the chiral and deconfinement transitions from a Curci-Ferrari-based Polyakov loop potential}",
    eprint = "2601.15839",
    archivePrefix = "arXiv",
    primaryClass = "hep-ph",
    month = "1",
    year = "2026"
}

@article{Duarte:2016iko,
    author = "Duarte, Anthony G. and Oliveira, Orlando and Silva, Paulo J.",
    title = "{Lattice Gluon and Ghost Propagators, and the Strong Coupling in Pure SU(3) Yang-Mills Theory: Finite Lattice Spacing and Volume Effects}",
    eprint = "1605.00594",
    archivePrefix = "arXiv",
    primaryClass = "hep-lat",
    doi = "10.1103/PhysRevD.94.014502",
    journal = "Phys. Rev. D",
    volume = "94",
    number = "1",
    pages = "014502",
    year = "2016"
}

@article{Bogolubsky:2009dc,
    author = "Bogolubsky, I. L. and Ilgenfritz, E. M. and Muller-Preussker, M. and Sternbeck, A.",
    title = "{Lattice gluodynamics computation of Landau gauge Green's functions in the deep infrared}",
    eprint = "0901.0736",
    archivePrefix = "arXiv",
    primaryClass = "hep-lat",
    reportNumber = "HU-EP-08-69, ADP-08-18-T678",
    doi = "10.1016/j.physletb.2009.04.076",
    journal = "Phys. Lett. B",
    volume = "676",
    pages = "69--73",
    year = "2009"
}

@article{Verschelde:2001ia,
    author = "Verschelde, H. and Knecht, K. and Van Acoleyen, K. and Vanderkelen, M.",
    title = "{The Nonperturbative groundstate of QCD and the local composite operator A(mu)**2}",
    eprint = "hep-th/0105018",
    archivePrefix = "arXiv",
    doi = "10.1016/S0370-2693(01)00929-7",
    journal = "Phys. Lett. B",
    volume = "516",
    pages = "307--313",
    year = "2001"
}

@article{Tissier:2011ey,
    author = "Tissier, Matthieu and Wschebor, Nicolas",
    title = "{An Infrared Safe perturbative approach to Yang-Mills correlators}",
    eprint = "1105.2475",
    archivePrefix = "arXiv",
    primaryClass = "hep-th",
    doi = "10.1103/PhysRevD.84.045018",
    journal = "Phys. Rev. D",
    volume = "84",
    pages = "045018",
    year = "2011"
}

@article{Aguilar:2023qqd,
    author = "Aguilar, A. C. and Ferreira, M. N. and Papavassiliou, J. and Santos, L. R.",
    title = "{Planar degeneracy of the three-gluon vertex}",
    eprint = "2305.05704",
    archivePrefix = "arXiv",
    primaryClass = "hep-ph",
    doi = "10.1140/epjc/s10052-023-11732-3",
    journal = "Eur. Phys. J. C",
    volume = "83",
    number = "6",
    pages = "549",
    year = "2023"
}

@article{Pinto-Gomez:2022brg,
    author = "Pinto-G{\'o}mez, F. and De Soto, F. and Ferreira, M. N. and Papavassiliou, J. and Rodr{\'\i}guez-Quintero, J.",
    title = "{Lattice three-gluon vertex in extended kinematics: Planar degeneracy}",
    eprint = "2208.01020",
    archivePrefix = "arXiv",
    primaryClass = "hep-ph",
    doi = "10.1016/j.physletb.2023.137737",
    journal = "Phys. Lett. B",
    volume = "838",
    pages = "137737",
    year = "2023"
}

@article{Pinto-Gomez:2024mrk,
    author = "Pinto-G{\'o}mez, F. and De Soto, F. and Rodr{\'\i}guez-Quintero, J.",
    title = "{Complete analysis of the Landau-gauge three-gluon vertex from lattice QCD}",
    eprint = "2404.08777",
    archivePrefix = "arXiv",
    primaryClass = "hep-ph",
    doi = "10.1103/PhysRevD.110.014005",
    journal = "Phys. Rev. D",
    volume = "110",
    number = "1",
    pages = "014005",
    year = "2024"
}

@article{Ball:1980ax,
    author = "Ball, James S. and Chiu, Ting-Wai",
    title = "{Analytic Properties of the Vertex Function in Gauge Theories. 2.}",
    reportNumber = "UU/HEP-80/3",
    doi = "10.1103/PhysRevD.22.2550",
    journal = "Phys. Rev. D",
    volume = "22",
    pages = "2550",
    year = "1980",
    note = "[Erratum: Phys.Rev.D 23, 3085 (1981)]"
}

@article{Pelaez:2013cpa,
    author = "Pelaez, Marcela and Tissier, Matthieu and Wschebor, Nicolas",
    title = "{Three-point correlation functions in Yang-Mills theory}",
    eprint = "1310.2594",
    archivePrefix = "arXiv",
    primaryClass = "hep-th",
    doi = "10.1103/PhysRevD.88.125003",
    journal = "Phys. Rev. D",
    volume = "88",
    pages = "125003",
    year = "2013"
}

@article{Pelaez:2021tpq,
    author = "Pel{\'a}ez, Marcela and Reinosa, Urko and Serreau, Julien and Tissier, Matthieu and Wschebor, Nicol{\'a}s",
    title = "{A window on infrared QCD with small expansion parameters}",
    eprint = "2106.04526",
    archivePrefix = "arXiv",
    primaryClass = "hep-th",
    doi = "10.1088/1361-6633/ac36b8",
    journal = "Rept. Prog. Phys.",
    volume = "84",
    number = "12",
    pages = "124202",
    year = "2021"
}

@article{Figueroa:2021sjm,
    author = "Figueroa, Felipe and Pel{\'a}ez, Marcela",
    title = "{One-loop unquenched three-gluon and ghost-gluon vertices in the Curci-Ferrari model}",
    eprint = "2110.09561",
    archivePrefix = "arXiv",
    primaryClass = "hep-th",
    doi = "10.1103/PhysRevD.105.094005",
    journal = "Phys. Rev. D",
    volume = "105",
    number = "9",
    pages = "094005",
    year = "2022"
}

@article{Barrios:2022hzr,
    author = "Barrios, Nahuel and Pel{\'a}ez, Marcela and Reinosa, Urko",
    title = "{Two-loop three-gluon vertex from the Curci-Ferrari model and its leading infrared behavior to all loop orders}",
    eprint = "2207.10704",
    archivePrefix = "arXiv",
    primaryClass = "hep-ph",
    doi = "10.1103/PhysRevD.106.114039",
    journal = "Phys. Rev. D",
    volume = "106",
    number = "11",
    pages = "114039",
    year = "2022"
}

@article{Smolyakov:1980wq,
    author = "Smolyakov, N. V.",
    title = "{Furry Theorem for Nonabelian Gauge Lagrangians}",
    reportNumber = "IFVE-81-3",
    doi = "10.1007/BF01016449",
    journal = "Theor. Math. Phys.",
    volume = "50",
    pages = "225--228",
    year = "1982"
}

@article{Aguilar:2019jsj,
    author = "Aguilar, A. C. and Ferreira, M. N. and Figueiredo, C. T. and Papavassiliou, J.",
    title = "{Nonperturbative Ball-Chiu construction of the three-gluon vertex}",
    eprint = "1903.01184",
    archivePrefix = "arXiv",
    primaryClass = "hep-ph",
    doi = "10.1103/PhysRevD.99.094010",
    journal = "Phys. Rev. D",
    volume = "99",
    number = "9",
    pages = "094010",
    year = "2019"
}

@article{Davydychev:1990jt,
    author = "Davydychev, Andrei I.",
    title = "{Some exact results for N point massive Feynman integrals}",
    reportNumber = "MGU-90-23-169",
    doi = "10.1063/1.529383",
    journal = "J. Math. Phys.",
    volume = "32",
    pages = "1052--1060",
    year = "1991"
}

@article{Davydychev:1990cq,
    author = "Davydychev, Andrei I.",
    title = "{General results for massive N point Feynman diagrams with different masses}",
    reportNumber = "MGU-90-56-202",
    doi = "10.1063/1.529914",
    journal = "J. Math. Phys.",
    volume = "33",
    pages = "358--369",
    year = "1992"
}

@article{Parrinello:1994wd,
    author = "Parrinello, C.",
    title = "{Exploratory study of the three gluon vertex on the lattice}",
    eprint = "hep-lat/9405024",
    archivePrefix = "arXiv",
    doi = "10.1103/PhysRevD.50.R4247",
    journal = "Phys. Rev. D",
    volume = "50",
    pages = "R4247--R4251",
    year = "1994"
}

@article{Cucchieri:2008qm,
    author = "Cucchieri, Attilio and Maas, Axel and Mendes, Tereza",
    title = "{Three-point vertices in Landau-gauge Yang-Mills theory}",
    eprint = "0803.1798",
    archivePrefix = "arXiv",
    primaryClass = "hep-lat",
    doi = "10.1103/PhysRevD.77.094510",
    journal = "Phys. Rev. D",
    volume = "77",
    pages = "094510",
    year = "2008"
}

@article{Cucchieri:2006tf,
    author = "Cucchieri, Attilio and Maas, Axel and Mendes, Tereza",
    title = "{Exploratory study of three-point Green's functions in Landau-gauge Yang-Mills theory}",
    eprint = "hep-lat/0605011",
    archivePrefix = "arXiv",
    doi = "10.1103/PhysRevD.74.014503",
    journal = "Phys. Rev. D",
    volume = "74",
    pages = "014503",
    year = "2006"
}

@article{Alles:1996ka,
    author = "Alles, B. and Henty, D. and Panagopoulos, H. and Parrinello, C. and Pittori, C. and Richards, D. G.",
    title = "{$\alpha_s$ from the nonperturbatively renormalised lattice three gluon vertex}",
    eprint = "hep-lat/9605033",
    archivePrefix = "arXiv",
    reportNumber = "FERMILAB-PUB-97-169-T, IFUP-TH-23-96, LPTHE-ORSAY-95-24, LTH-367",
    doi = "10.1016/S0550-3213(97)00483-5",
    journal = "Nucl. Phys. B",
    volume = "502",
    pages = "325--342",
    year = "1997"
}

@article{Vujinovic:2018nqc,
    author = "Vujinovic, Milan and Mendes, Tereza",
    title = "{Probing the tensor structure of lattice three-gluon vertex in Landau gauge}",
    eprint = "1807.03673",
    archivePrefix = "arXiv",
    primaryClass = "hep-lat",
    doi = "10.1103/PhysRevD.99.034501",
    journal = "Phys. Rev. D",
    volume = "99",
    number = "3",
    pages = "034501",
    year = "2019"
}

@article{Duarte:2016ieu,
    author = "Duarte, Anthony G. and Oliveira, Orlando and Silva, Paulo J.",
    title = "{Further Evidence For Zero Crossing On The Three Gluon Vertex}",
    eprint = "1607.03831",
    archivePrefix = "arXiv",
    primaryClass = "hep-lat",
    doi = "10.1103/PhysRevD.94.074502",
    journal = "Phys. Rev. D",
    volume = "94",
    number = "7",
    pages = "074502",
    year = "2016"
}

@article{Boucaud:2017obn,
    author = "Boucaud, Ph. and De Soto, F. and Rodr{\'\i}guez-Quintero, J. and Zafeiropoulos, S.",
    title = "{Refining the detection of the zero crossing for the three-gluon vertex in symmetric and asymmetric momentum subtraction schemes}",
    eprint = "1701.07390",
    archivePrefix = "arXiv",
    primaryClass = "hep-lat",
    doi = "10.1103/PhysRevD.95.114503",
    journal = "Phys. Rev. D",
    volume = "95",
    number = "11",
    pages = "114503",
    year = "2017"
}

@article{Athenodorou:2016oyh,
    author = "Athenodorou, A. and Binosi, D. and Boucaud, Ph. and De Soto, F. and Papavassiliou, J. and Rodriguez-Quintero, J. and Zafeiropoulos, S.",
    title = "{On the zero crossing of the three-gluon vertex}",
    eprint = "1607.01278",
    archivePrefix = "arXiv",
    primaryClass = "hep-ph",
    doi = "10.1016/j.physletb.2016.08.065",
    journal = "Phys. Lett. B",
    volume = "761",
    pages = "444--449",
    year = "2016"
}

@article{Eichmann:2014xya,
    author = "Eichmann, Gernot and Williams, Richard and Alkofer, Reinhard and Vujinovic, Milan",
    title = "{Three-gluon vertex in Landau gauge}",
    eprint = "1402.1365",
    archivePrefix = "arXiv",
    primaryClass = "hep-ph",
    doi = "10.1103/PhysRevD.89.105014",
    journal = "Phys. Rev. D",
    volume = "89",
    number = "10",
    pages = "105014",
    year = "2014"
}

@article{Aguilar:2023mdv,
    author = "Aguilar, A. C. and Ferreira, M. N. and Oliveira, B. M. and Papavassiliou, J. and Santos, L. R.",
    title = "{Schwinger poles of the three-gluon vertex: symmetry and dynamics}",
    eprint = "2306.16283",
    archivePrefix = "arXiv",
    primaryClass = "hep-ph",
    doi = "10.1140/epjc/s10052-023-12058-w",
    journal = "Eur. Phys. J. C",
    volume = "83",
    number = "10",
    pages = "889",
    year = "2023"
}

@article{Aguilar:2022thg,
    author = "Aguilar, A. C. and De Soto, F. and Ferreira, M. N. and Papavassiliou, J. and Pinto-G{\'o}mez, F. and Roberts, C. D. and Rodr{\'\i}guez-Quintero, J.",
    title = "{Schwinger mechanism for gluons from lattice QCD}",
    eprint = "2211.12594",
    archivePrefix = "arXiv",
    primaryClass = "hep-ph",
    reportNumber = "NJU-INP 068/22",
    doi = "10.1016/j.physletb.2023.137906",
    journal = "Phys. Lett. B",
    volume = "841",
    pages = "137906",
    year = "2023"
}

@article{Aguilar:2021lke,
    author = "Aguilar, A. C. and De Soto, F. and Ferreira, M. N. and Papavassiliou, J. and Rodr{\'\i}guez-Quintero, J.",
    title = "{Infrared facets of the three-gluon vertex}",
    eprint = "2102.04959",
    archivePrefix = "arXiv",
    primaryClass = "hep-ph",
    doi = "10.1016/j.physletb.2021.136352",
    journal = "Phys. Lett. B",
    volume = "818",
    pages = "136352",
    year = "2021"
}

@article{Huber:2020keu,
    author = "Huber, Markus Q.",
    title = "{Correlation functions of Landau gauge Yang-Mills theory}",
    eprint = "2003.13703",
    archivePrefix = "arXiv",
    primaryClass = "hep-ph",
    doi = "10.1103/PhysRevD.101.114009",
    journal = "Phys. Rev. D",
    volume = "101",
    pages = "114009",
    year = "2020"
}

@article{Aguilar:2019uob,
    author = "Aguilar, A. C. and De Soto, F. and Ferreira, M. N. and Papavassiliou, J. and Rodr{\'\i}guez-Quintero, J. and Zafeiropoulos, S.",
    title = "{Gluon propagator and three-gluon vertex with dynamical quarks}",
    eprint = "1912.12086",
    archivePrefix = "arXiv",
    primaryClass = "hep-ph",
    doi = "10.1140/epjc/s10052-020-7741-0",
    journal = "Eur. Phys. J. C",
    volume = "80",
    number = "2",
    pages = "154",
    year = "2020"
}

@article{Huber:2018ned,
    author = "Huber, Markus Q.",
    title = "{Nonperturbative properties of Yang{\textendash}Mills theories}",
    eprint = "1808.05227",
    archivePrefix = "arXiv",
    primaryClass = "hep-ph",
    doi = "10.1016/j.physrep.2020.04.004",
    journal = "Phys. Rept.",
    volume = "879",
    pages = "1--92",
    year = "2020"
}

@article{Binosi:2017rwj,
    author = "Binosi, Daniele and Papavassiliou, Joannis",
    title = "{Coupled dynamics in gluon mass generation and the impact of the three-gluon vertex}",
    eprint = "1709.09964",
    archivePrefix = "arXiv",
    primaryClass = "hep-ph",
    doi = "10.1103/PhysRevD.97.054029",
    journal = "Phys. Rev. D",
    volume = "97",
    number = "5",
    pages = "054029",
    year = "2018"
}

@article{Cyrol:2016tym,
    author = "Cyrol, Anton K. and Fister, Leonard and Mitter, Mario and Pawlowski, Jan M. and Strodthoff, Nils",
    title = "{Landau gauge Yang-Mills correlation functions}",
    eprint = "1605.01856",
    archivePrefix = "arXiv",
    primaryClass = "hep-ph",
    doi = "10.1103/PhysRevD.94.054005",
    journal = "Phys. Rev. D",
    volume = "94",
    number = "5",
    pages = "054005",
    year = "2016"
}

@article{Huber:2016tvc,
    author = "Huber, Markus Q.",
    title = "{Correlation functions of three-dimensional Yang-Mills theory from Dyson-Schwinger equations}",
    eprint = "1602.02038",
    archivePrefix = "arXiv",
    primaryClass = "hep-th",
    doi = "10.1103/PhysRevD.93.085033",
    journal = "Phys. Rev. D",
    volume = "93",
    number = "8",
    pages = "085033",
    year = "2016"
}

@article{Huber:2014isa,
    author = "Huber, Markus Q. and Campagnari, Davide R. and Reinhardt, Hugo",
    title = "{Vertex functions of Coulomb gauge Yang--Mills theory}",
    eprint = "1410.4766",
    archivePrefix = "arXiv",
    primaryClass = "hep-ph",
    doi = "10.1103/PhysRevD.91.025014",
    journal = "Phys. Rev. D",
    volume = "91",
    number = "2",
    pages = "025014",
    year = "2015"
}

@article{Eichmann:2021zuv,
    author = "Eichmann, Gernot and Pawlowski, Jan M. and Silva, Jo{\~a}o M.",
    title = "{Mass generation in Landau-gauge Yang-Mills theory}",
    eprint = "2107.05352",
    archivePrefix = "arXiv",
    primaryClass = "hep-ph",
    doi = "10.1103/PhysRevD.104.114016",
    journal = "Phys. Rev. D",
    volume = "104",
    number = "11",
    pages = "114016",
    year = "2021"
}

@article{deBoer:1995dh,
    author = "de Boer, Jan and Skenderis, Kostas and van Nieuwenhuizen, Peter and Waldron, Andrew",
    title = "{On the renormalizability and unitarity of the Curci-Ferrari model for massive vector bosons}",
    eprint = "hep-th/9510167",
    archivePrefix = "arXiv",
    reportNumber = "ITP-SB-95-43",
    doi = "10.1016/0370-2693(95)01455-1",
    journal = "Phys. Lett. B",
    volume = "367",
    pages = "175--182",
    year = "1996"
}

@article{Gracey:2002yt,
    author = "Gracey, J. A.",
    title = "{Three loop MS-bar renormalization of the Curci-Ferrari model and the dimension two BRST invariant composite operator in QCD}",
    eprint = "hep-th/0211144",
    archivePrefix = "arXiv",
    reportNumber = "LTH-565",
    doi = "10.1016/S0370-2693(02)03077-0",
    journal = "Phys. Lett. B",
    volume = "552",
    pages = "101--110",
    year = "2003"
}

@article{Tissier:2008nw,
    author = "Tissier, Matthieu and Wschebor, Nicolas",
    title = "{Gauged supersymmetries in Yang-Mills theory}",
    eprint = "0809.1880",
    archivePrefix = "arXiv",
    primaryClass = "hep-th",
    doi = "10.1103/PhysRevD.79.065008",
    journal = "Phys. Rev. D",
    volume = "79",
    pages = "065008",
    year = "2009"
}

@article{Dudal:2002pq,
    author = "Dudal, D. and Verschelde, H. and Sorella, S. P.",
    title = "{The Anomalous dimension of the composite operator A**2 in the Landau gauge}",
    eprint = "hep-th/0212182",
    archivePrefix = "arXiv",
    doi = "10.1016/S0370-2693(03)00043-1",
    journal = "Phys. Lett. B",
    volume = "555",
    pages = "126--131",
    year = "2003"
}

@article{Tissier:2010ts,
    author = "Tissier, Matthieu and Wschebor, Nicolas",
    title = "{Infrared propagators of Yang-Mills theory from perturbation theory}",
    eprint = "1004.1607",
    archivePrefix = "arXiv",
    primaryClass = "hep-ph",
    doi = "10.1103/PhysRevD.82.101701",
    journal = "Phys. Rev. D",
    volume = "82",
    pages = "101701",
    year = "2010"
}

@article{Taylor:1971ff,
    author = "Taylor, J. C.",
    title = "{Ward Identities and Charge Renormalization of the Yang-Mills Field}",
    doi = "10.1016/0550-3213(71)90297-5",
    journal = "Nucl. Phys. B",
    volume = "33",
    pages = "436--444",
    year = "1971"
}

@article{Wschebor:2007vh,
    author = "Wschebor, Nicolas",
    title = "{Some non-renormalization theorems in Curci-Ferrari model}",
    eprint = "hep-th/0701127",
    archivePrefix = "arXiv",
    doi = "10.1142/S0217751X08040469",
    journal = "Int. J. Mod. Phys. A",
    volume = "23",
    pages = "2961--2973",
    year = "2008"
}

@article{Delduc:1989uc,
    author = "Delduc, F. and Sorella, S. P.",
    title = "{A Note on Some Nonlinear Covariant Gauges in {Yang-Mills} Theory}",
    reportNumber = "CERN-TH-5514/89",
    doi = "10.1016/0370-2693(89)90684-9",
    journal = "Phys. Lett. B",
    volume = "231",
    pages = "408--410",
    year = "1989"
}

@article{Curci:1976bt,
    author = "Curci, G. and Ferrari, R.",
    title = "{On a Class of Lagrangian Models for Massive and Massless Yang-Mills Fields}",
    doi = "10.1007/BF02729999",
    journal = "Nuovo Cim. A",
    volume = "32",
    pages = "151--168",
    year = "1976"
}

@article{Corell:2018yil,
    author = "Corell, Lukas and Cyrol, Anton K. and Mitter, Mario and Pawlowski, Jan M. and Strodthoff, Nils",
    title = "{Correlation functions of three-dimensional Yang-Mills theory from the FRG}",
    eprint = "1803.10092",
    archivePrefix = "arXiv",
    primaryClass = "hep-ph",
    doi = "10.21468/SciPostPhys.5.6.066",
    journal = "SciPost Phys.",
    volume = "5",
    number = "6",
    pages = "066",
    year = "2018"
}

@article{EuropeanTwistedMass:2008pab,
    author = "Alexandrou, C. and others",
    collaboration = "European Twisted Mass",
    title = "{Light baryon masses with dynamical twisted mass fermions}",
    eprint = "0803.3190",
    archivePrefix = "arXiv",
    primaryClass = "hep-lat",
    reportNumber = "LPT-ORSAY-08-32, IRFU-08-29, DESY-08-032, SFB-CPP-08-19, ROM2F-2008-06",
    doi = "10.1103/PhysRevD.78.014509",
    journal = "Phys. Rev. D",
    volume = "78",
    pages = "014509",
    year = "2008"
}

@article{Fodor:2012gf,
    author = "Fodor, Zoltan and Hoelbling, Christian",
    title = "{Light Hadron Masses from Lattice QCD}",
    eprint = "1203.4789",
    archivePrefix = "arXiv",
    primaryClass = "hep-lat",
    reportNumber = "WUB-11-12",
    doi = "10.1103/RevModPhys.84.449",
    journal = "Rev. Mod. Phys.",
    volume = "84",
    pages = "449",
    year = "2012"
}

@article{EuropeanTwistedMass:2014osg,
    author = "Carrasco, N. and others",
    collaboration = "European Twisted Mass",
    title = "{Up, down, strange and charm quark masses with N$_f$ = 2+1+1 twisted mass lattice QCD}",
    eprint = "1403.4504",
    archivePrefix = "arXiv",
    primaryClass = "hep-lat",
    reportNumber = "PREPRINT-MITP-14-020, ROM2F-2014-01, RM3-TH-14-4",
    doi = "10.1016/j.nuclphysb.2014.07.025",
    journal = "Nucl. Phys. B",
    volume = "887",
    pages = "19--68",
    year = "2014"
}

@article{Bussone:2023kag,
    author = "Bussone, Andrea and Conigli, Alessandro and Frison, Julien and Herdo{\'\i}za, Gregorio and Pena, Carlos and Preti, David and S{\'a}ez, Alejandro and Ugarrio, Javier",
    collaboration = "Alpha",
    title = "{Hadronic physics from a Wilson fermion mixed-action approach: charm quark mass and $D_{(s)}$ meson decay constants}",
    eprint = "2309.14154",
    archivePrefix = "arXiv",
    primaryClass = "hep-lat",
    reportNumber = "IFT-UAM/CSIC-23-114",
    doi = "10.1140/epjc/s10052-024-12816-4",
    journal = "Eur. Phys. J. C",
    volume = "84",
    number = "5",
    pages = "506",
    year = "2024"
}

@article{Bazavov:2017dsy,
    author = "Bazavov, A. and Petreczky, P. and Weber, J. H.",
    title = "{Equation of State in 2+1 Flavor QCD at High Temperatures}",
    eprint = "1710.05024",
    archivePrefix = "arXiv",
    primaryClass = "hep-lat",
    doi = "10.1103/PhysRevD.97.014510",
    journal = "Phys. Rev. D",
    volume = "97",
    number = "1",
    pages = "014510",
    year = "2018"
}

@article{Mandula:1987rh,
    author = "Mandula, J. E. and Ogilvie, M.",
    title = "{The Gluon Is Massive: A Lattice Calculation of the Gluon Propagator in the Landau Gauge}",
    doi = "10.1016/0370-2693(87)91541-3",
    journal = "Phys. Lett. B",
    volume = "185",
    pages = "127--132",
    year = "1987"
}

@article{Catumba:2021yly,
    author = "Catumba, Guilherme T. R. and Oliveira, Orlando and Silva, Paulo J.",
    title = "{Another look at the Landau gauge three-gluon vertex}",
    eprint = "2111.10312",
    archivePrefix = "arXiv",
    primaryClass = "hep-lat",
    doi = "10.1051/epjconf/202225802008",
    journal = "EPJ Web Conf.",
    volume = "258",
    pages = "02008",
    year = "2022"
}

@article{Bonnet:2000kw,
    author = "Bonnet, Frederic D. R. and Bowman, Patrick O. and Leinweber, Derek B. and Williams, Anthony G.",
    title = "{Infrared behavior of the gluon propagator on a large volume lattice}",
    eprint = "hep-lat/0002020",
    archivePrefix = "arXiv",
    reportNumber = "ADP-00-10-T396",
    doi = "10.1103/PhysRevD.62.051501",
    journal = "Phys. Rev. D",
    volume = "62",
    pages = "051501",
    year = "2000"
}

@article{Maas:2020zjp,
    author = "Maas, Axel and Vujinovi{\'c}, Milan",
    title = "{More on the three-gluon vertex in SU(2) Yang-Mills theory in three and four dimensions}",
    eprint = "2006.08248",
    archivePrefix = "arXiv",
    primaryClass = "hep-lat",
    doi = "10.21468/SciPostPhysCore.5.2.019",
    journal = "SciPost Phys. Core",
    volume = "5",
    pages = "019",
    year = "2022"
}

@article{Kizilersu:2021jen,
    author = {K{\i}z{\i}lers{\"u}, Ay{\c{s}}e and Oliveira, Orlando and Silva, Paulo J. and Skullerud, Jon-Ivar and Sternbeck, Andr{\'e}},
    title = "{Quark-gluon vertex from Nf=2 lattice QCD}",
    eprint = "2103.02945",
    archivePrefix = "arXiv",
    primaryClass = "hep-lat",
    reportNumber = "ADP-20-32/T1142",
    doi = "10.1103/PhysRevD.103.114515",
    journal = "Phys. Rev. D",
    volume = "103",
    number = "11",
    pages = "114515",
    year = "2021"
}

@article{Silva:2010vx,
    author = "Silva, Paulo J. and Oliveira, Orlando",
    editor = "Rossi, Giancarlo",
    title = "{Unquenching the Landau Gauge Lattice Propagators and the Gribov Problem}",
    eprint = "1011.0483",
    archivePrefix = "arXiv",
    primaryClass = "hep-lat",
    doi = "10.22323/1.105.0287",
    journal = "PoS",
    volume = "LATTICE2010",
    pages = "287",
    year = "2010"
}

@article{Bowman:2004jm,
    author = "Bowman, Patrick O. and Heller, Urs M. and Leinweber, Derek B. and Parappilly, Maria B. and Williams, Anthony G.",
    title = "{Unquenched gluon propagator in Landau gauge}",
    eprint = "hep-lat/0402032",
    archivePrefix = "arXiv",
    reportNumber = "ADP-03-142-T577",
    doi = "10.1103/PhysRevD.70.034509",
    journal = "Phys. Rev. D",
    volume = "70",
    pages = "034509",
    year = "2004"
}

@article{Marques:2023cmi,
    author = "Marques, Jesuel and Kalusche, Gerhard and Mendes, Tereza and Silva, Paulo J. and Skullerud, Jon-Ivar and Oliveira, Orlando",
    title = "{The quark propagator and quark-gluon vertex from lattice QCD at finite temperature}",
    eprint = "2301.10607",
    archivePrefix = "arXiv",
    primaryClass = "hep-lat",
    reportNumber = "PoS(LATTICE2022)280",
    doi = "10.22323/1.430.0280",
    journal = "PoS",
    volume = "LATTICE2022",
    pages = "280",
    year = "2023"
}

@article{Colaco:2024gmt,
    author = "Cola{\c{c}}o, Manuel and Oliveira, Orlando and Silva, Paulo J.",
    title = "{Four-gluon vertex from lattice QCD}",
    eprint = "2401.12008",
    archivePrefix = "arXiv",
    primaryClass = "hep-lat",
    doi = "10.1103/PhysRevD.109.074502",
    journal = "Phys. Rev. D",
    volume = "109",
    number = "7",
    pages = "074502",
    year = "2024"
}

@article{Oliveira:2012eh,
    author = "Oliveira, Orlando and Silva, Paulo J.",
    title = "{The lattice Landau gauge gluon propagator: lattice spacing and volume dependence}",
    eprint = "1207.3029",
    archivePrefix = "arXiv",
    primaryClass = "hep-lat",
    doi = "10.1103/PhysRevD.86.114513",
    journal = "Phys. Rev. D",
    volume = "86",
    pages = "114513",
    year = "2012"
}

@article{Cucchieri:2007rg,
    author = "Cucchieri, A. and Mendes, T.",
    title = "{Constraints on the IR behavior of the gluon propagator in Yang-Mills theories}",
    eprint = "0712.3517",
    archivePrefix = "arXiv",
    primaryClass = "hep-lat",
    doi = "10.1103/PhysRevLett.100.241601",
    journal = "Phys. Rev. Lett.",
    volume = "100",
    pages = "241601",
    year = "2008"
}

@article{Iritani:2009mp,
    author = "Iritani, Takumi and Suganuma, Hideo and Iida, Hideaki",
    title = "{Gluon-propagator functional form in the Landau gauge in SU(3) lattice QCD: Yukawa-type gluon propagator and anomalous gluon spectral function}",
    eprint = "0908.1311",
    archivePrefix = "arXiv",
    primaryClass = "hep-lat",
    doi = "10.1103/PhysRevD.80.114505",
    journal = "Phys. Rev. D",
    volume = "80",
    pages = "114505",
    year = "2009"
}

@article{Bornyakov:2009ug,
    author = "Bornyakov, V. G. and Mitrjushkin, V. K. and Muller-Preussker, M.",
    title = "{SU(2) lattice gluon propagator: Continuum limit, finite-volume effects and infrared mass scale m(IR)}",
    eprint = "0912.4475",
    archivePrefix = "arXiv",
    primaryClass = "hep-lat",
    reportNumber = "HU-EP-09-63, ITEP-LAT-2009-18",
    doi = "10.1103/PhysRevD.81.054503",
    journal = "Phys. Rev. D",
    volume = "81",
    pages = "054503",
    year = "2010"
}

@article{vonSmekal:1997ohs,
    author = "von Smekal, Lorenz and Alkofer, Reinhard and Hauck, Andreas",
    title = "{The Infrared behavior of gluon and ghost propagators in Landau gauge QCD}",
    eprint = "hep-ph/9705242",
    archivePrefix = "arXiv",
    reportNumber = "ANL-PHY-8683-TH-97, UNITU-THEP-7-1997, TU-GK-97-002",
    doi = "10.1103/PhysRevLett.79.3591",
    journal = "Phys. Rev. Lett.",
    volume = "79",
    pages = "3591--3594",
    year = "1997"
}

@article{Alkofer:2000wg,
    author = "Alkofer, Reinhard and von Smekal, Lorenz",
    title = "{The Infrared behavior of QCD Green's functions: Confinement dynamical symmetry breaking, and hadrons as relativistic bound states}",
    eprint = "hep-ph/0007355",
    archivePrefix = "arXiv",
    reportNumber = "UNITUE-THEP-00-09, FAU-TP3-00-8",
    doi = "10.1016/S0370-1573(01)00010-2",
    journal = "Phys. Rept.",
    volume = "353",
    pages = "281",
    year = "2001"
}

@article{Fischer:2002hna,
    author = "Fischer, C. S. and Alkofer, Reinhard",
    title = "{Infrared exponents and running coupling of SU(N) Yang-Mills theories}",
    eprint = "hep-ph/0202202",
    archivePrefix = "arXiv",
    reportNumber = "UNITUE-THEP-07-2002",
    doi = "10.1016/S0370-2693(02)01809-9",
    journal = "Phys. Lett. B",
    volume = "536",
    pages = "177--184",
    year = "2002"
}

@article{Zwanziger:2001kw,
    author = "Zwanziger, Daniel",
    title = "{Nonperturbative Landau gauge and infrared critical exponents in QCD}",
    eprint = "hep-th/0109224",
    archivePrefix = "arXiv",
    doi = "10.1103/PhysRevD.65.094039",
    journal = "Phys. Rev. D",
    volume = "65",
    pages = "094039",
    year = "2002"
}

@article{Maas:2004se,
    author = "Maas, Axel and Wambach, Jochen and Gruter, Burghard and Alkofer, Reinhard",
    title = "{High-temperature limit of Landau-gauge Yang-Mills theory}",
    eprint = "hep-ph/0408074",
    archivePrefix = "arXiv",
    doi = "10.1140/epjc/s2004-02004-3",
    journal = "Eur. Phys. J. C",
    volume = "37",
    pages = "335--357",
    year = "2004"
}

@article{Aguilar:2007ie,
    author = "Aguilar, Arlene C. and Papavassiliou, Joannis",
    title = "{Power-law running of the effective gluon mass}",
    eprint = "0708.4320",
    archivePrefix = "arXiv",
    primaryClass = "hep-ph",
    doi = "10.1140/epja/i2008-10535-4",
    journal = "Eur. Phys. J. A",
    volume = "35",
    pages = "189--205",
    year = "2008"
}

@article{Aguilar:2008xm,
    author = "Aguilar, A. C. and Binosi, D. and Papavassiliou, J.",
    title = "{Gluon and ghost propagators in the Landau gauge: Deriving lattice results from Schwinger-Dyson equations}",
    eprint = "0802.1870",
    archivePrefix = "arXiv",
    primaryClass = "hep-ph",
    reportNumber = "ECT*-08-02",
    doi = "10.1103/PhysRevD.78.025010",
    journal = "Phys. Rev. D",
    volume = "78",
    pages = "025010",
    year = "2008"
}

@article{Boucaud:2008ky,
    author = "Boucaud, Philippe and Leroy, J. P. and Le Yaouanc, A. and Micheli, J. and Pene, O. and Rodriguez-Quintero, J.",
    title = "{On the IR behaviour of the Landau-gauge ghost propagator}",
    eprint = "0803.2161",
    archivePrefix = "arXiv",
    primaryClass = "hep-ph",
    reportNumber = "UHU-FP-08-010, LPT-ORSAY-08-28",
    doi = "10.1088/1126-6708/2008/06/099",
    journal = "JHEP",
    volume = "06",
    pages = "099",
    year = "2008"
}

@article{Boucaud:2008ji,
    author = "Boucaud, Philippe and Leroy, J-P. and Yaouanc, A. Le and Micheli, J. and Pene, O. and Rodriguez-Quintero, J.",
    title = "{IR finiteness of the ghost dressing function from numerical resolution of the ghost SD equation}",
    eprint = "0801.2721",
    archivePrefix = "arXiv",
    primaryClass = "hep-ph",
    reportNumber = "LPT-ORSAY-08-44, UHU-FT-0803",
    doi = "10.1088/1126-6708/2008/06/012",
    journal = "JHEP",
    volume = "06",
    pages = "012",
    year = "2008"
}

@article{Huber:2007kc,
    author = "Huber, Markus Q. and Alkofer, Reinhard and Fischer, Christian S. and Schwenzer, Kai",
    title = "{The Infrared behavior of Landau gauge Yang-Mills theory in d=2, d=3 and d=4 dimensions}",
    eprint = "0705.3809",
    archivePrefix = "arXiv",
    primaryClass = "hep-ph",
    doi = "10.1016/j.physletb.2007.10.073",
    journal = "Phys. Lett. B",
    volume = "659",
    pages = "434--440",
    year = "2008"
}

@article{Huber:2012kd,
    author = "Huber, Markus Q. and von Smekal, Lorenz",
    title = "{On the influence of three-point functions on the propagators of Landau gauge Yang-Mills theory}",
    eprint = "1211.6092",
    archivePrefix = "arXiv",
    primaryClass = "hep-th",
    doi = "10.1007/JHEP04(2013)149",
    journal = "JHEP",
    volume = "04",
    pages = "149",
    year = "2013"
}

@article{Fischer:2020xnb,
    author = "Fischer, Christian S. and Huber, Markus Q.",
    title = "{Landau gauge Yang-Mills propagators in the complex momentum plane}",
    eprint = "2007.11505",
    archivePrefix = "arXiv",
    primaryClass = "hep-ph",
    doi = "10.1103/PhysRevD.102.094005",
    journal = "Phys. Rev. D",
    volume = "102",
    number = "9",
    pages = "094005",
    year = "2020"
}

@article{Fischer:2008uz,
    author = "Fischer, Christian S. and Maas, Axel and Pawlowski, Jan M.",
    title = "{On the infrared behavior of Landau gauge Yang-Mills theory}",
    eprint = "0810.1987",
    archivePrefix = "arXiv",
    primaryClass = "hep-ph",
    doi = "10.1016/j.aop.2009.07.009",
    journal = "Annals Phys.",
    volume = "324",
    pages = "2408--2437",
    year = "2009"
}

@article{Ellwanger:1995qf,
    author = "Ellwanger, Ulrich and Hirsch, Manfred and Weber, Axel",
    title = "{Flow equations for the relevant part of the pure Yang-Mills action}",
    eprint = "hep-th/9506019",
    archivePrefix = "arXiv",
    reportNumber = "LPTHE-ORSAY-95-39",
    doi = "10.1007/s002880050073",
    journal = "Z. Phys. C",
    volume = "69",
    pages = "687--698",
    year = "1996"
}

@article{Pawlowski:2003hq,
    author = "Pawlowski, Jan M. and Litim, Daniel F. and Nedelko, Sergei and von Smekal, Lorenz",
    title = "{Infrared behavior and fixed points in Landau gauge QCD}",
    eprint = "hep-th/0312324",
    archivePrefix = "arXiv",
    reportNumber = "FAU-TP3-03-12, CERN-TH-2003-300, SHEP-03-33",
    doi = "10.1103/PhysRevLett.93.152002",
    journal = "Phys. Rev. Lett.",
    volume = "93",
    pages = "152002",
    year = "2004"
}

@article{Fischer:2004uk,
    author = "Fischer, Christian S. and Gies, Holger",
    title = "{Renormalization flow of Yang-Mills propagators}",
    eprint = "hep-ph/0408089",
    archivePrefix = "arXiv",
    reportNumber = "HD-THEP-04-29, IPPP-04-17, DCPT-04-34",
    doi = "10.1088/1126-6708/2004/10/048",
    journal = "JHEP",
    volume = "10",
    pages = "048",
    year = "2004"
}

@article{Fischer:2006vf,
    author = "Fischer, Christian S. and Pawlowski, Jan M.",
    title = "{Uniqueness of infrared asymptotics in Landau gauge Yang-Mills theory}",
    eprint = "hep-th/0609009",
    archivePrefix = "arXiv",
    reportNumber = "HD-THEP-06-20",
    doi = "10.1103/PhysRevD.75.025012",
    journal = "Phys. Rev. D",
    volume = "75",
    pages = "025012",
    year = "2007"
}

@article{Dupuis:2020fhh,
    author = "Dupuis, N. and Canet, L. and Eichhorn, A. and Metzner, W. and Pawlowski, J. M. and Tissier, M. and Wschebor, N.",
    title = "{The nonperturbative functional renormalization group and its applications}",
    eprint = "2006.04853",
    archivePrefix = "arXiv",
    primaryClass = "cond-mat.stat-mech",
    doi = "10.1016/j.physrep.2021.01.001",
    journal = "Phys. Rept.",
    volume = "910",
    pages = "1--114",
    year = "2021"
}

@article{Maas:2007uv,
    author = "Maas, Axel",
    title = "{Two and three-point Green's functions in two-dimensional Landau-gauge Yang-Mills theory}",
    eprint = "0704.0722",
    archivePrefix = "arXiv",
    primaryClass = "hep-lat",
    doi = "10.1103/PhysRevD.75.116004",
    journal = "Phys. Rev. D",
    volume = "75",
    pages = "116004",
    year = "2007"
}

@article{Maas:2011se,
    author = "Maas, Axel",
    title = "{Describing gauge bosons at zero and finite temperature}",
    eprint = "1106.3942",
    archivePrefix = "arXiv",
    primaryClass = "hep-ph",
    doi = "10.1016/j.physrep.2012.11.002",
    journal = "Phys. Rept.",
    volume = "524",
    pages = "203--300",
    year = "2013"
}

@article{Boucaud:2011ug,
    author = "Boucaud, Ph. and Leroy, J. P. and Yaouanc, A. Le and Micheli, J. and Pene, O. and Rodriguez-Quintero, J.",
    title = "{The Infrared Behaviour of the Pure Yang-Mills Green Functions}",
    eprint = "1109.1936",
    archivePrefix = "arXiv",
    primaryClass = "hep-ph",
    doi = "10.1007/s00601-011-0301-2",
    journal = "Few Body Syst.",
    volume = "53",
    pages = "387--436",
    year = "2012"
}

@article{Faddeev:1967fc,
    author = "Faddeev, L. D. and Popov, V. N.",
    editor = "Hsu, Jong-Ping and Fine, D.",
    title = "{Feynman Diagrams for the Yang-Mills Field}",
    doi = "10.1016/0370-2693(67)90067-6",
    journal = "Phys. Lett. B",
    volume = "25",
    pages = "29--30",
    year = "1967"
}

@article{Gribov:1977wm,
    author = "Gribov, V. N.",
    editor = "Nyiri, J.",
    title = "{Quantization of Nonabelian Gauge Theories}",
    reportNumber = "LENINGRAD-77-367",
    doi = "10.1016/0550-3213(78)90175-X",
    journal = "Nucl. Phys. B",
    volume = "139",
    pages = "1",
    year = "1978"
}

@article{Politzer:1973fx,
    author = "Politzer, H. David",
    editor = "Taylor, J. C.",
    title = "{Reliable Perturbative Results for Strong Interactions?}",
    doi = "10.1103/PhysRevLett.30.1346",
    journal = "Phys. Rev. Lett.",
    volume = "30",
    pages = "1346--1349",
    year = "1973"
}

@article{Gross:1973id,
    author = "Gross, David J. and Wilczek, Frank",
    editor = "Taylor, J. C.",
    title = "{Ultraviolet Behavior of Nonabelian Gauge Theories}",
    doi = "10.1103/PhysRevLett.30.1343",
    journal = "Phys. Rev. Lett.",
    volume = "30",
    pages = "1343--1346",
    year = "1973"
}

@article{Vandersickel:2012tz,
    author = "Vandersickel, N. and Zwanziger, Daniel",
    title = "{The Gribov problem and QCD dynamics}",
    eprint = "1202.1491",
    archivePrefix = "arXiv",
    primaryClass = "hep-th",
    doi = "10.1016/j.physrep.2012.07.003",
    journal = "Phys. Rept.",
    volume = "520",
    pages = "175--251",
    year = "2012"
}

@article{Zwanziger:1988jt,
    author = "Zwanziger, Daniel",
    title = "{Action From the Gribov Horizon}",
    reportNumber = "Print-88-0822 (NEW YORK)",
    doi = "10.1016/0550-3213(89)90263-0",
    journal = "Nucl. Phys. B",
    volume = "321",
    pages = "591--604",
    year = "1989"
}

@article{Zwanziger:1989mf,
    author = "Zwanziger, D.",
    title = "{Local and Renormalizable Action From the Gribov Horizon}",
    doi = "10.1016/0550-3213(89)90122-3",
    journal = "Nucl. Phys. B",
    volume = "323",
    pages = "513--544",
    year = "1989"
}

@article{Dudal:2008sp,
    author = "Dudal, David and Gracey, John A. and Sorella, Silvio Paolo and Vandersickel, Nele and Verschelde, Henri",
    title = "{A Refinement of the Gribov-Zwanziger approach in the Landau gauge: Infrared propagators in harmony with the lattice results}",
    eprint = "0806.4348",
    archivePrefix = "arXiv",
    primaryClass = "hep-th",
    reportNumber = "LTH-789",
    doi = "10.1103/PhysRevD.78.065047",
    journal = "Phys. Rev. D",
    volume = "78",
    pages = "065047",
    year = "2008"
}

@article{Serreau:2012cg,
    author = "Serreau, J. and Tissier, M.",
    title = "{Lifting the Gribov ambiguity in Yang-Mills theories}",
    eprint = "1202.3432",
    archivePrefix = "arXiv",
    primaryClass = "hep-th",
    doi = "10.1016/j.physletb.2012.04.041",
    journal = "Phys. Lett. B",
    volume = "712",
    pages = "97--103",
    year = "2012"
}

@article{Barrios:2021cks,
    author = "Barrios, Nahuel and Gracey, John A. and Pel{\'a}ez, Marcela and Reinosa, Urko",
    title = "{Two-loop corrections to the QCD propagators within the Curci-Ferrari model}",
    eprint = "2103.16218",
    archivePrefix = "arXiv",
    primaryClass = "hep-th",
    doi = "10.1103/PhysRevD.104.094019",
    journal = "Phys. Rev. D",
    volume = "104",
    number = "9",
    pages = "094019",
    year = "2021"
}

@article{Barrios:2020ubx,
    author = "Barrios, Nahuel and Pel{\'a}ez, Marcela and Reinosa, Urko and Wschebor, Nicol{\'a}s",
    title = "{The ghost-antighost-gluon vertex from the Curci-Ferrari model: Two-loop corrections}",
    eprint = "2009.00875",
    archivePrefix = "arXiv",
    primaryClass = "hep-th",
    doi = "10.1103/PhysRevD.102.114016",
    journal = "Phys. Rev. D",
    volume = "102",
    pages = "114016",
    year = "2020"
}

@article{Barrios:2024ixj,
    author = "Barrios, Nahuel and De Fabritiis, Philipe and Pel{\'a}ez, Marcela",
    title = "{Four-gluon vertex from the Curci-Ferrari model at one-loop order}",
    eprint = "2403.17056",
    archivePrefix = "arXiv",
    primaryClass = "hep-th",
    doi = "10.1103/PhysRevD.109.L091502",
    journal = "Phys. Rev. D",
    volume = "109",
    number = "9",
    pages = "L091502",
    year = "2024"
}

@article{Gracey:2019xom,
    author = "Gracey, John A. and Pel{\'a}ez, Marcela and Reinosa, Urko and Tissier, Matthieu",
    title = "{Two loop calculation of Yang-Mills propagators in the Curci-Ferrari model}",
    eprint = "1905.07262",
    archivePrefix = "arXiv",
    primaryClass = "hep-th",
    doi = "10.1103/PhysRevD.100.034023",
    journal = "Phys. Rev. D",
    volume = "100",
    number = "3",
    pages = "034023",
    year = "2019"
}

@article{Pelaez:2015tba,
    author = "Pel{\'a}ez, Marcela and Tissier, Matthieu and Wschebor, Nicol{\'a}s",
    title = "{Quark-gluon vertex from the Landau gauge Curci-Ferrari model}",
    eprint = "1504.05157",
    archivePrefix = "arXiv",
    primaryClass = "hep-th",
    doi = "10.1103/PhysRevD.92.045012",
    journal = "Phys. Rev. D",
    volume = "92",
    number = "4",
    pages = "045012",
    year = "2015"
}

@article{Pelaez:2014mxa,
    author = "Pel{\'a}ez, M. and Tissier, M. and Wschebor, N.",
    title = "{Two-point correlation functions of QCD in the Landau gauge}",
    eprint = "1407.2005",
    archivePrefix = "arXiv",
    primaryClass = "hep-th",
    doi = "10.1103/PhysRevD.90.065031",
    journal = "Phys. Rev. D",
    volume = "90",
    pages = "065031",
    year = "2014"
}

@article{Pelaez:2017bhh,
    author = "Pel{\'a}ez, Marcela and Reinosa, Urko and Serreau, Julien and Tissier, Matthieu and Wschebor, Nicol{\'a}s",
    title = "{Small parameters in infrared quantum chromodynamics}",
    eprint = "1703.10288",
    archivePrefix = "arXiv",
    primaryClass = "hep-th",
    doi = "10.1103/PhysRevD.96.114011",
    journal = "Phys. Rev. D",
    volume = "96",
    number = "11",
    pages = "114011",
    year = "2017"
}

@article{Aguilar:2019kxz,
    author = "Aguilar, A. C. and Ferreira, M. N. and Figueiredo, C. T. and Papavassiliou, J.",
    title = "{Gluon mass scale through nonlinearities and vertex interplay}",
    eprint = "1909.09826",
    archivePrefix = "arXiv",
    primaryClass = "hep-ph",
    doi = "10.1103/PhysRevD.100.094039",
    journal = "Phys. Rev. D",
    volume = "100",
    number = "9",
    pages = "094039",
    year = "2019"
}

@article{Ferreira:2025anh,
    author = "Ferreira, M. N. and Papavassiliou, J.",
    title = "{Gluon mass scale through the Schwinger mechanism}",
    eprint = "2501.01080",
    archivePrefix = "arXiv",
    primaryClass = "hep-ph",
    doi = "10.1016/j.ppnp.2025.104186",
    journal = "Prog. Part. Nucl. Phys.",
    volume = "144",
    pages = "104186",
    year = "2025"
}

@article{Ferreira:2024czk,
    author = "Ferreira, M. N. and Papavassiliou, J.",
    title = "{Nonlinear Schwinger mechanism in QCD, and Fredholm alternatives theorem}",
    eprint = "2407.04392",
    archivePrefix = "arXiv",
    primaryClass = "hep-ph",
    doi = "10.1140/epjc/s10052-024-13223-5",
    journal = "Eur. Phys. J. C",
    volume = "84",
    number = "8",
    pages = "835",
    year = "2024"
}

@article{Reinosa:2017qtf,
    author = "Reinosa, Urko and Serreau, Julien and Tissier, Matthieu and Wschebor, Nicol{\'a}s",
    title = "{How nonperturbative is the infrared regime of Landau gauge Yang-Mills correlators?}",
    eprint = "1703.04041",
    archivePrefix = "arXiv",
    primaryClass = "hep-th",
    doi = "10.1103/PhysRevD.96.014005",
    journal = "Phys. Rev. D",
    volume = "96",
    number = "1",
    pages = "014005",
    year = "2017"
}

@article{Siringo:2015wtx,
    author = "Siringo, Fabio",
    title = "{Analytical study of Yang{\textendash}Mills theory in the infrared from first principles}",
    eprint = "1511.01015",
    archivePrefix = "arXiv",
    primaryClass = "hep-ph",
    doi = "10.1016/j.nuclphysb.2016.04.028",
    journal = "Nucl. Phys. B",
    volume = "907",
    pages = "572--596",
    year = "2016"
}

@article{Siringo:2016jrc,
    author = "Siringo, Fabio",
    title = "{Analytic structure of QCD propagators in Minkowski space}",
    eprint = "1605.07357",
    archivePrefix = "arXiv",
    primaryClass = "hep-ph",
    doi = "10.1103/PhysRevD.94.114036",
    journal = "Phys. Rev. D",
    volume = "94",
    number = "11",
    pages = "114036",
    year = "2016"
}

@article{Comitini:2021kxj,
    author = "Comitini, Giorgio and Rizzo, Daniele and Battello, Massimiliano and Siringo, Fabio",
    title = "{Screened massive expansion of the quark propagator in the Landau gauge}",
    eprint = "2108.00417",
    archivePrefix = "arXiv",
    primaryClass = "hep-th",
    doi = "10.1103/PhysRevD.104.074020",
    journal = "Phys. Rev. D",
    volume = "104",
    number = "7",
    pages = "074020",
    year = "2021"
}

@article{Bopsin:2025vhz,
    author = "Bopsin, Gustavo B. and El-Bennich, Bruno and Krein, Gast{\~a}o and Serna, Fernando E. and da Silveira, Roberto C.",
    title = "{Parton distribution and fragmentation functions with massive gluons}",
    eprint = "2507.12544",
    archivePrefix = "arXiv",
    primaryClass = "hep-ph",
    doi = "10.1103/ygrj-pglg",
    journal = "Phys. Rev. D",
    volume = "112",
    number = "11",
    pages = "114023",
    year = "2025"
}

@article{Song:2019qoh,
    author = "Song, Yifan and Baym, Gordon and Hatsuda, Tetsuo and Kojo, Toru",
    title = "{Effective repulsion in dense quark matter from nonperturbative gluon exchange}",
    eprint = "1905.01005",
    archivePrefix = "arXiv",
    primaryClass = "astro-ph.HE",
    reportNumber = "RIKEN-iTHEMS-Report-19",
    doi = "10.1103/PhysRevD.100.034018",
    journal = "Phys. Rev. D",
    volume = "100",
    number = "3",
    pages = "034018",
    year = "2019"
}

@article{Suenaga:2019jjv,
    author = "Suenaga, Daiki and Kojo, Toru",
    title = "{Gluon propagator in two-color dense QCD: Massive Yang-Mills approach at one-loop}",
    eprint = "1905.08751",
    archivePrefix = "arXiv",
    primaryClass = "hep-ph",
    doi = "10.1103/PhysRevD.100.076017",
    journal = "Phys. Rev. D",
    volume = "100",
    number = "7",
    pages = "076017",
    year = "2019"
}

@article{Kojo:2021knn,
    author = "Kojo, Toru and Suenaga, Daiki",
    title = "{Thermal quarks and gluon propagators in two-color dense QCD}",
    eprint = "2102.07231",
    archivePrefix = "arXiv",
    primaryClass = "hep-ph",
    doi = "10.1103/PhysRevD.103.094008",
    journal = "Phys. Rev. D",
    volume = "103",
    number = "9",
    pages = "094008",
    year = "2021"
}

@misc{supplementalMaterial,
   author = {Daniel O. R. Azevedo and Nahuel Barrios and Marcela Pel\'{a}ez and Nicol\'{a}s Wschebor}, 
    year = {2026}, 
    title = {Mathematica notebook of "Three-gluon vertex planar degeneracy: a perturbative explanation": vertex expressions}, 
    doi = {https://doi.org/10.60895/redata/QPYRCA},
}

\end{document}